\documentclass{aa}
\usepackage{graphicx}
\usepackage{natbib}
\input epsf.sty

\def \grss {GRS 1915+105~}
\def \grs {GRS 1915+105}

\def \sax {{\it Beppo}SAX}

\begin{document}
\title{The complex time behaviour of the microquasar GRS~1915+105 in the $\rho$ 
class observed with \sax. III: The hard X-ray delay and limit cycle mapping}
\author{F.~Massa\inst{1}
\and E.~Massaro\inst{2}
\and T.~Mineo\inst{3}
\and A.~D'A\`i\inst{4}
\and M.~Feroci\inst{5}
\and P.~Casella\inst{6}
\and T.~Belloni\inst{7}
\institute{INFN, Sezione Roma 1,Piazzale A. Moro 2, I-00185 Roma, Italy
\and Dipartimento di Fisica, Universit\`a La Sapienza, Piazzale A. Moro 2, I-00185 Roma, Italy
\and INAF, IASF, Sez. di Palermo, via U. La Malfa 153, I-90146 Palermo, Italy
\and Dipartimento di Fisica, Universit\`a di Palermo, Via Archirafi 36, I-90123 Palermo, Italy
\and INAF, IAPS, Sez. di Roma, via del Fosso del Cavaliere 100, I-00113 Roma, Italy
\and INAF, Osservatorio Astronomico di Roma, I-23807 Monte Porzio Catone, Italy
\and INAF, Osservatorio Astronomico di Brera, via E. Bianchi 46, I-23807 Merate, 
Italy
}}
\offprints{mineo@iasf-palermo.inaf.it}
\date{Received ....; accepted ....}

\markboth{F. Massa et al.: The complex time behaviour of GRS~1915+105}
{F. Massa et al.: The complex time behaviour of GRS~1915+105}

\abstract{The microquasar GRS1915+105 was observed by BeppoSAX in October 2000
for about ten days while the source was in $\rho$ mode, 
which is characterized by a quasi-regular type I bursting activity. }
{This paper presents a systematic analysis of the delay of the hard and soft X-ray emission 
at the burst peaks. 
The lag, also apparent from the comparison of the [1.7-3.4] keV light curves with those in 
the [6.8-10.2] keV range, is evaluated and studied as a function of time, spectral parameters,
and flux.}
{We apply the limit cycle mapping technique, using as independent variables
the count rate and the mean photon rate. 
The results using this technique were also cross-checked using a more standard approach 
with the cross-correlation methods.
Data are organized in runs, each relative to a continuous observation interval.}
{The detected hard-soft delay changes in the course of the pointing from $\sim$3 s to $\sim$10 
s and presents a clear correlation with the baseline count rate.}
{}
\keywords{stars: binaries: close - stars: individual: GRS 1915+105 -
 X-rays: stars - black hole physics}

\authorrunning{F. Massa et al.}
\titlerunning{The complex behaviour of GRS~1915+105 in the $\rho$ class. III: 
The hard X-ray delay}

\maketitle


\section{Introduction}
\label{section1}

The various phenomena occurring in accretion disks around black 
holes (BH) can exhibit complex patterns likely originated by non-linear processes. 
In the case of the microquasar \grs, such complex processes produce a large variety 
of behaviours, ranging from a rather steady and noisy emission to the occurrence of long 
recurrent burst series.
In two previous papers on \grs, we investigated the time evolution of the burst 
series properties in the so-called $\rho$ class \citep{bel_2}  using the data 
collected in the course of a long pointing of \sax~ in October 2000 
\citep[][hereafter Paper I]{mass_1}; the results of the spectral analysis 
were described in \citet[][Paper II]{mineo_1}.

\begin{figure*}[ht!]
\center{
\includegraphics[height=8.cm,angle=0,scale=1.0]{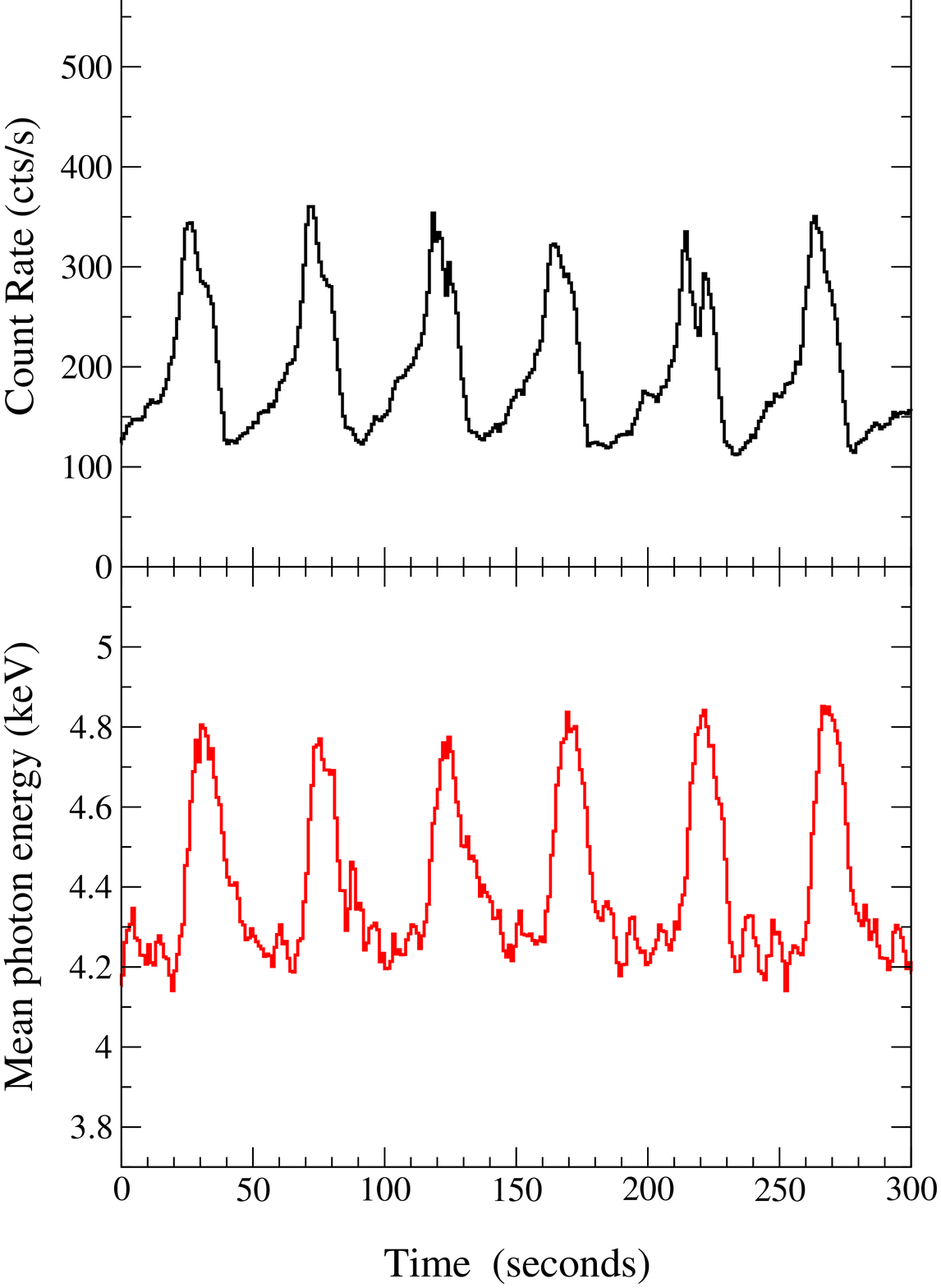}
\includegraphics[height=8.cm,angle=0,scale=1.0]{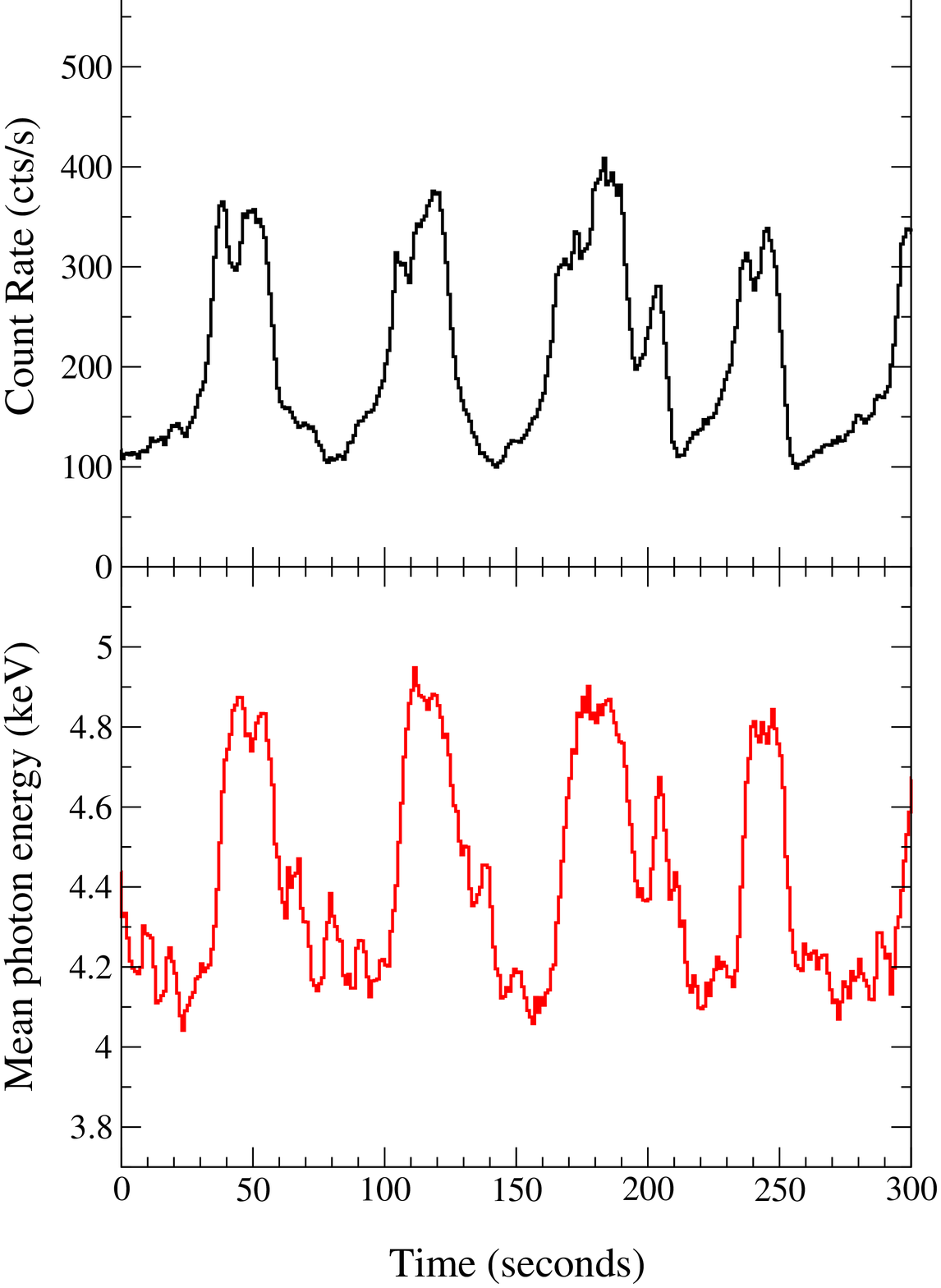}
\includegraphics[height=8.cm,angle=0,scale=1.0]{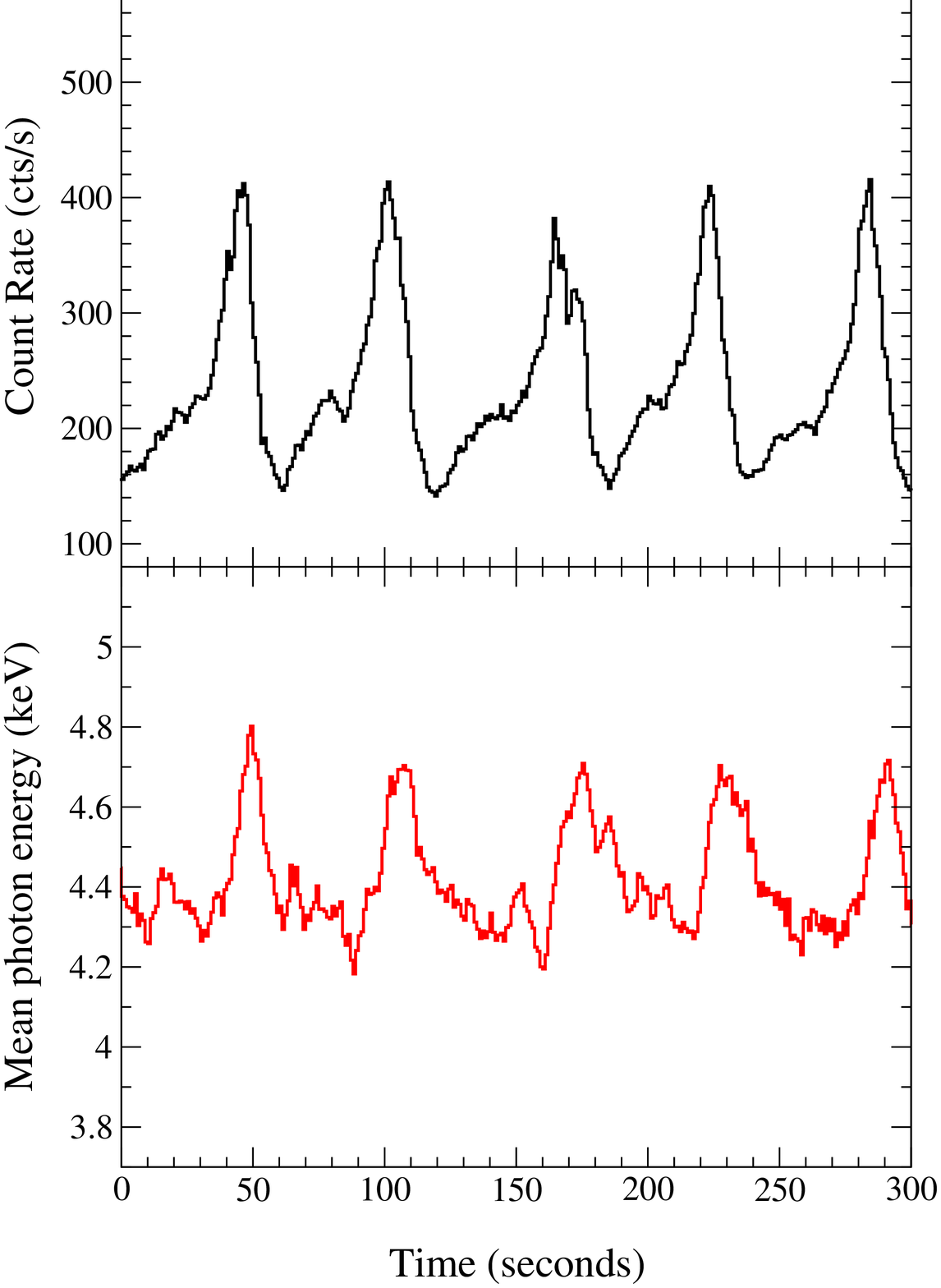}
}
\caption[]{
Two 300-second long segments of the count rate (upper panels) and mean photon 
energy (lower panels) curves of the MECS [1.7 - 10] keV data series A8b (left),
E5 (centre), and F7 (right) after a running average smoothing over five bins. 
The bin size of all series is 1 s.
}
\label{fig1}
\end{figure*}

In these two papers, we introduced the $regular$ and $irregular$ modes of the $\rho$ 
class on the basis of the stability of the recurrence time of bursts $T_{rec}$, 
derived from Fourier and wavelet spectra, and on the multiplicity of pulses.  
We introduced a synthetic nomenclature useful for classifying the various types
of bursting. 
Fourier periodograms of individual series were classified in three types, namely S, T, 
and M, according to the occurrence of a single, two, or many prominent peaks, respectively.
We defined three stability classes, which were denoted as 0, 1, and 2 and corresponded to
a decrease of the fluctuations of the highest power time-scale in the wavelet 
scalograms.
Thus series of the S2 type correspond to the most regular ones, whereas the M0 ones are the most 
irregular.
The entire pointing was divided into three successive intervals identified by a Roman 
numeral: I (from the start time of the observation to 1.7$\times$10${^5}$ s), II 
(from this time to 3.8$\times$10${^5}$ s), and III (from 4.0$\times$10${^5}$ s to 
6.0$\times$10${^5}$ s). 
In the interval II, \grss was in the $irregular$ mode, while in the two others it was
mainly in the $regular$ one.
We observed an increase in the baseline count rate from interval II to interval III; 
the fractional final increase was $\sim$ 18\% and took place in $\sim$ 20 ks.

In Paper II, we described the results of a time-resolved spectral analysis that was 
performed splitting the entire burst cycle into a few segments:
the {\it Slow Leading Trail} ({\it SLT}) from the minimum level to about the half 
height of the {\it Pulse} ({\it P}) and followed by the {\it Fast Decaying Tail}, or 
{\it FDT}, in which the count rate decreases to its minimum. 
The first two segments were moreover separated into two parts of equal duration, named 
{\it SLT-1}, {\it SLT-2} and {\it P-1}, {\it P-2}, respectively. 
Bursts are superimposed onto a baseline level ({\it BL}), which showed a step-like increase
from interval II to III (Paper I).
For clarity's sake, we report in Table~\ref{summary} a log of the acronyms and labels used in Paper 
I and II.


\begin{table*}
\label{summary}
\caption{Acronyms and labels used in Paper I and II.}
\begin{center}
\begin{tabular}{ll}
\hline\hline
\multicolumn{2}{l}{Time intervals of the \sax~ pointing.}  \\
\\
I   & 0--170 ks   characterised by regular bursts (the series A8b is an example) \\
II  & 170--380 ks characterised by irregular bursts (the series E5 is an example) \\
     & 380--400 ks characterised by a count rate increase\\
III & 400--600 ks characterised by regular bursts (the series F7 is an example)\\
\\
\multicolumn{2}{l}{Characterisation of individual series
by Fourier periodograms. }  \\
\\
S & single prominent peak \\
T & two  prominent peaks  \\
M & many prominent peaks  \\
\\
\multicolumn{2}{l}{Stability class according to the fluctuations' level 
  in wavelet scalograms.} \\
\\
0 & high  fluctuations \\
1 & medium  fluctuations \\
2 & low fluctuations \\
\\
\multicolumn{2}{l}{Burst segments used in the Paper II spectral analysis.}  \\
\\
$SLT$-1  &  first part of the {\it Slow Leading Trail} \\
$SLT$-2  &  second part of the {\it Slow Leading Trail} \\
$P$-1    &  first part of the {\it Pulse} \\
$P$-2    &  second part of the {\it Pulse} \\
$FDT$    &{\it Fast Decaying Tail}
\\
\hline
\end{tabular}
\end{center}
\end{table*}

Lags between the hard and soft X-ray emission, hereafter named 
{\it hard-X delay} (HXD), have been observed in \grss and in
other X-ray binaries \citep{utt_1,cass_1}.
The HXD occurrence was reported since early observations of the $\rho$ class 
in terms of temperature evolution of the multi-temperature disk black body in the 
course of the burst  \citep{taa_1, pau_1}.
Similar changes were also found in the spectral analysis presented in Paper II:
the disk $kT$ changed from about 1.1 keV in the {\it SLT} to about 1.6 keV in the 
first segment of the {\it Pulse} ({\it P1}), reaching values above 2 keV in the 
second segment ({\it P2}) and decreasing to {\it P1} values during the {\it FDT}.
The same temperature evolution was reported by \citet{neil_2} using RXTE observations.
On time-scales shorter than 1 s, a hard phase lag of the high-energy component with
respect to the low-energy one was observed in \grss and other accreting black 
hole candidates on the basis of Fourier transforms. 
\citet{cui_1} reported for \grss a hard X-ray phase lag associated with the 67 Hz 
quasi-periodic oscillation and showed the existence of both hard and soft phaselags 
up to $\sim$ 1 s.
\citet{reig_1} found that a lag was also present in QPOs at lower frequencies 
[0.6--8] Hz and that their amplitude is correlated with the frequency, while 
\citet{muno_1} indicated the existence of both positive and negative lags, 
according to X-ray and radio intensity levels; in particular, the phase lag sign 
changed from positive to negative as radio emission increased.
\citet{jan_2} considered the same {\it RXTE} observations of \grss of 
\citet{bel_2} and, applying a cross-correlation 
to the energy-selected light curves in the [1.5$-$6] keV and [6.4$-$14.6] keV bands,
found a delay of $\sim$1 s only in the $\rho$ and $\kappa$ classes.
More recently, \citet{neil_2} in their analysis of $\rho$ class data defined
the HXD on the basis of the phase separation between the maxima on folded burst profiles
in the two energy ranges [2$-$5] and [12$-$45] keV.
They found that the phase lags depend on burst multiplicity: double-peaked bursts 
have significantly longer lags ($\Delta \phi$=0.08$\pm$0.05) than  single-peaked 
ones ($\Delta \phi$=0.02$\pm$0.02); we note that the latter result is 
compatible with a zero lag at one standard deviation. 
For a typical recurrence time of their data ($\sim$64 s), the former lag translates 
into a time distance of $\sim$4.6 s. 

The analysis presented in this paper is devoted to the correlated spectral-timing 
variations of  \grss emission in the $\rho$ class; more specifically, we focus 
on the analysis of the HXD.
We developed specific tools for the study of the spectral-timing behaviour by means of a limit cycle 
mapping and applied these methods to the investigation of both $regular$ and $irregular$ 
data series to investigate how the HXD changes across the long BeppoSAX observation.
After a brief description of data and the HXD phenomenon, we describe the methods developed
for mapping limit cycle behaviour in a suitable parameter space and for estimating the HXD.
Then our results are discussed and compared with previous findings.
We note that our approach and results will also be useful when comparing the $\rho$ class 
observed in \grss with other sources showing similar behaviour, and possibly the 
same underlying physical mechanisms as the newly discovered IGR J17091–3624 
 \citep{alt_1}, which exhibits very similar variability patterns.

\begin{figure}[h]
 \vspace{-1.5cm}
\includegraphics[height=8.0cm,angle=0,scale=1.0]{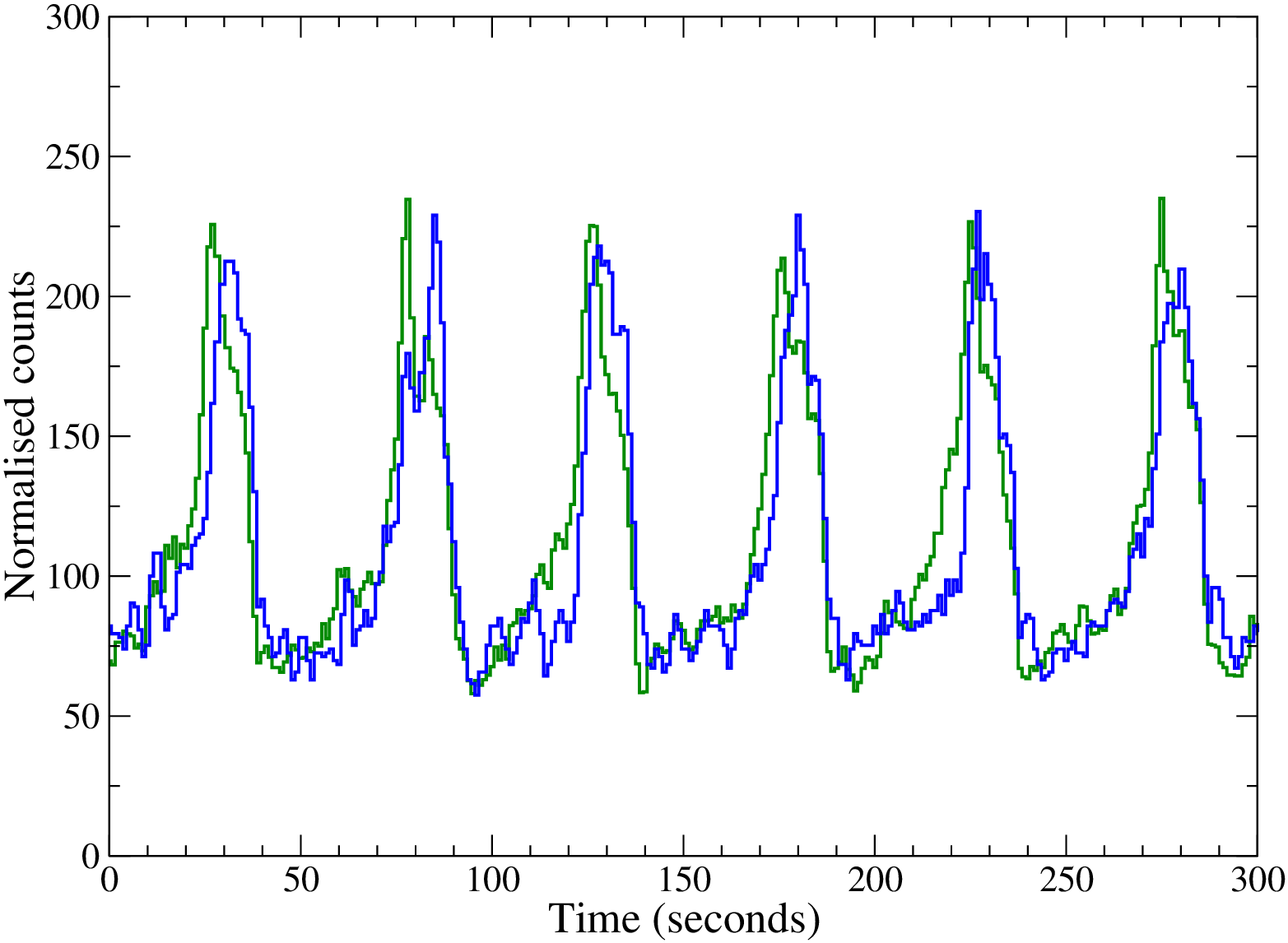} \\
 \vspace{-1.5cm}
\includegraphics[height=8.0cm,angle=0,scale=1.0]{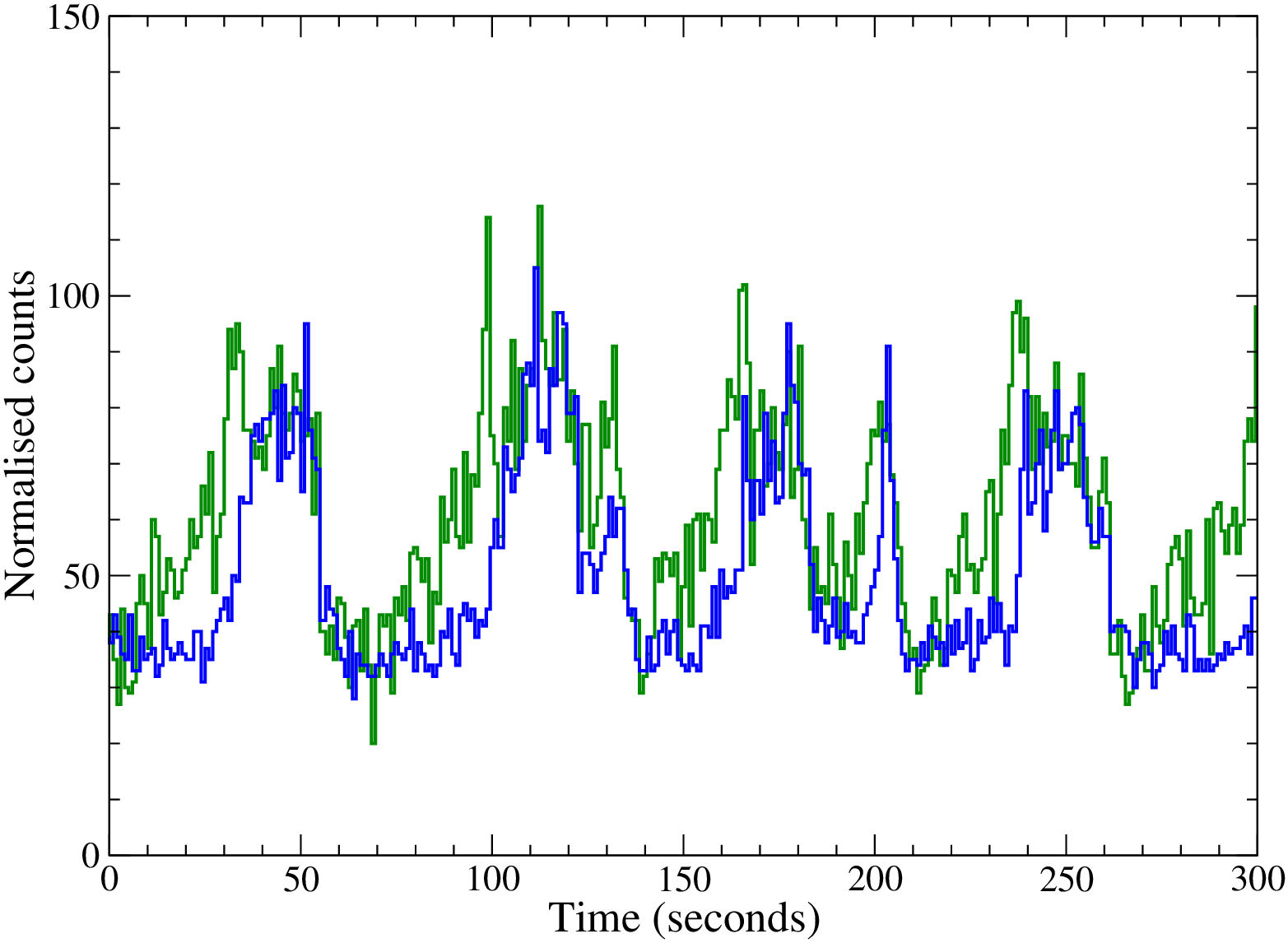} \\
 \vspace{-1.5cm}
\includegraphics[height=8.0cm,angle=0,scale=1.0]{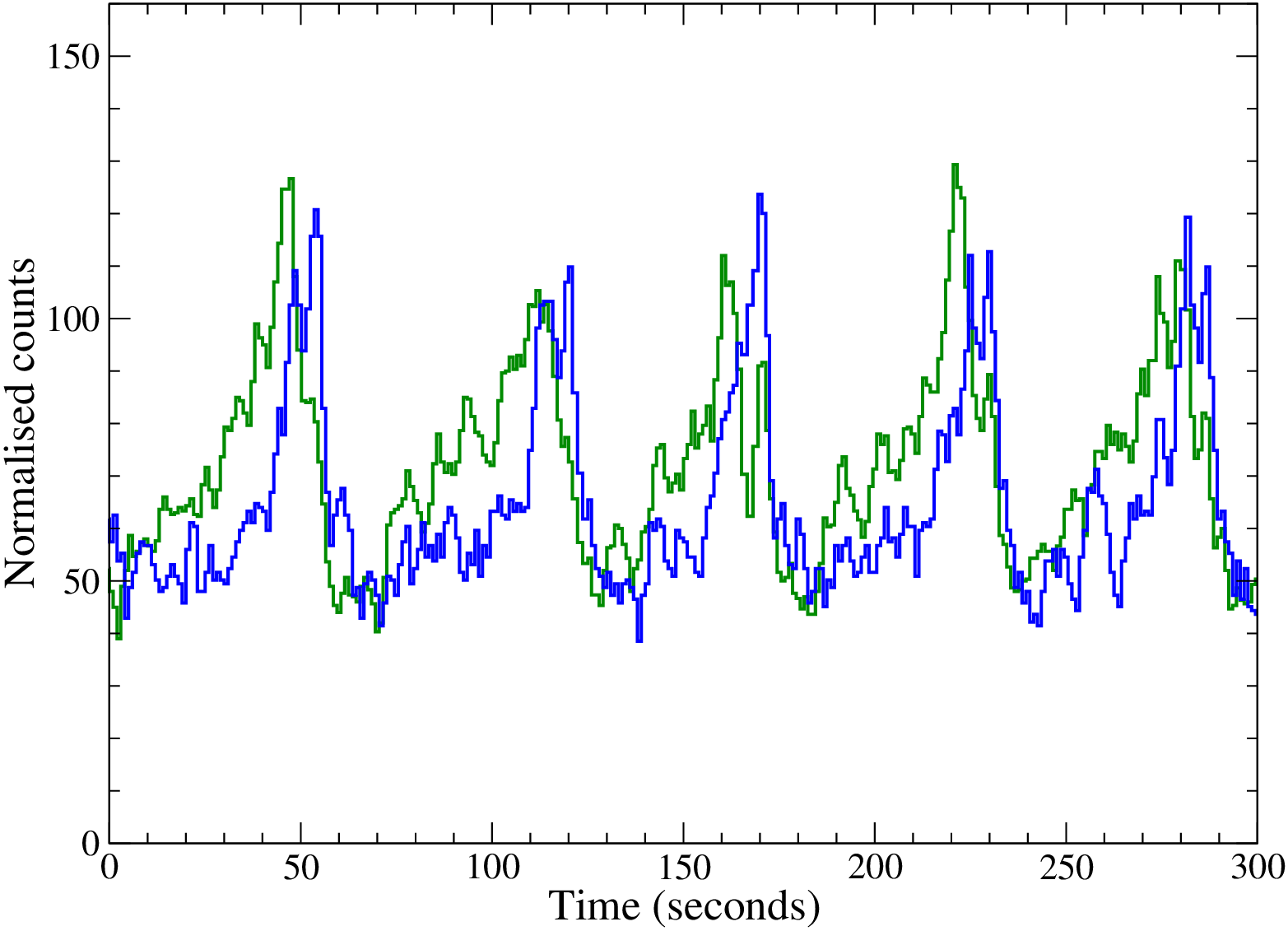}
\caption[]{
Portions of MECS light curves in the energy ranges [1.3 - 3.4] keV,
green curve, and [6.8 - 10.2] keV, blue curve. From top to bottom,
examples from time interval I (time series A8b), time interval II (E5),
and time interval III (F7). 
The time bin is 1 s.
}
\label{fig2}
\end{figure} 

\section{Observations and Data Reduction}
\label{section2}

General descriptions of the long \sax~ observation of \grss performed in 
October 2000 are given in Papers I and II, and we refer to those papers for more 
details.
This observation started on October 20 (MJD 51837.894) and terminated on October 
29 after an overall duration of 768.79 ks.
In this paper, our analysis is performed only on data obtained with the Medium 
Energy Concentrator Spectrometer \citep[MECS,][]{boe_1} in the energy range 
[1.7--10] keV, whose statistics are high enough to provide significant information 
on individual bursts. 
As shown in Paper I, data at higher energy, as those obtained with the PDS instrument,
are noisier and burst features cannot always be clearly established.   
Data are organized in runs, each relative to a continuous observation interval, and 
named with letters and sequential numbers. 
We use the same 52 data series that were already considered in Paper II; they are all within the first 
600.725 s of the pointing when the source remained almost stable in the 
$\rho$ class. The main parameters of the considered runs are given in Table \ref{table1}.

In addition to the count rate time series, we also consider in the following analysis   
the series describing the evolution of the mean energy of photons, which are good for 
properly describing the recurring bursting behaviour of the source in a two-dimensional 
space. 
These series were obtained by  computing the arithmetic mean 
$\langle C \rangle $ of the energy channels assigned to each photon in every time 
bin and then converting the result with the instrumental gain relationship

\begin{equation}
\langle E \rangle = 0.0232 + 0.0464 \langle C \rangle ~~~~~{\rm keV.}
\end{equation}
\noindent

Three examples of count rate and energy series, which are extracted from the light 
curves A8b (S2 type, $regular$), E5 (M0 type, $irregular$) and F7 (T2 type, $regular$), 
the same already used in Paper I as representative of the different modes of the $\rho$ 
class, are shown in Fig.~\ref{fig1}.
Considering the high scatter of the data due either to statistical fluctuations 
or to an intrinsic high-frequency variability, we smoothed the high-frequency noise 
by means of a running average filter over a time window of five bins.
Curves of the mean photon energy are generally limited in the quite 
narrow range between 4.2 keV and 4.8 keV, but their variations are very 
apparent.
Considering that the typical root-mean-squared (rms) dispersion in each bin 
is close and generally lower than 0.1 keV, these variations are significant and consistent 
with the spectral evolution presented in Paper II.
In our analysis the considered data series had the time bin width of one second.

\section{The hard X-ray delay}
\label{section3}

The occurrence of a delay of a few seconds between the emission at energies 
above 6 keV with respect to that at lower energies is clearly apparent from the 
short segments of the data series A8b, E5, and F7, shown in the panels of Fig.~\ref{fig2}, 
where data in the two energy ranges [1.7 -- 3.4] keV and [6.8 -- 10.2] keV
are compared. 
For the sake of clarity, high-energy count rates were scaled to achieve values
comparable to those at lower energies.
The count rate increase of {\it SLT} in the low-energy band anticipates
the one in the high-energy band, and peaks within the {\it Pulse} do not appear synchronous, 
whereas a lag at the end of the {\it FDT} is clearly resolved.
In Paper I, we already mentioned this effect and reported that it appears even
larger when comparing MECS with PDS data at energies higher than 15 keV. 

\begin{figure}[h!]
\includegraphics[height=8.5cm,angle=-90,scale=1.0]{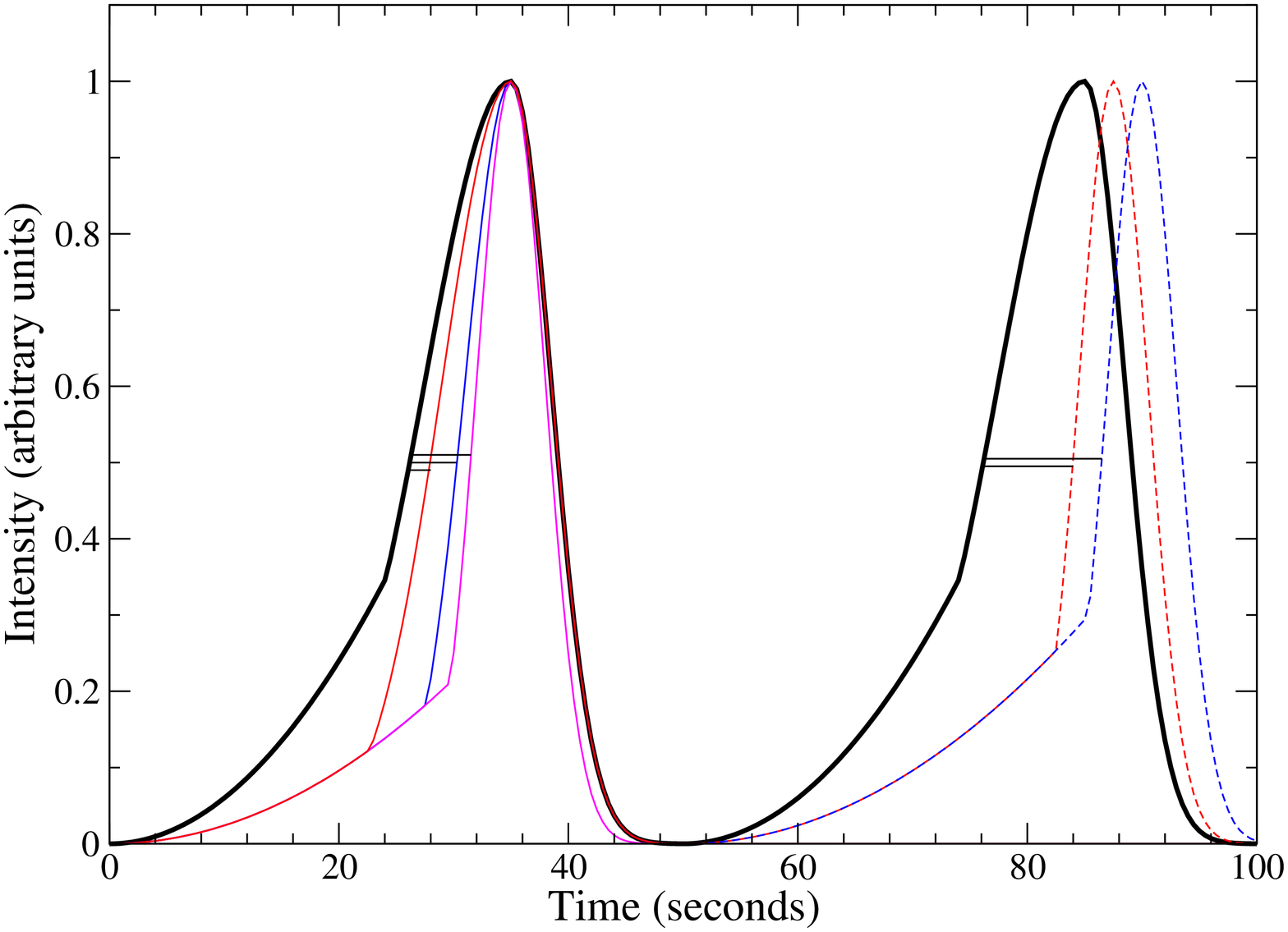}
\includegraphics[height=8.5cm,angle=-90,scale=1.0]{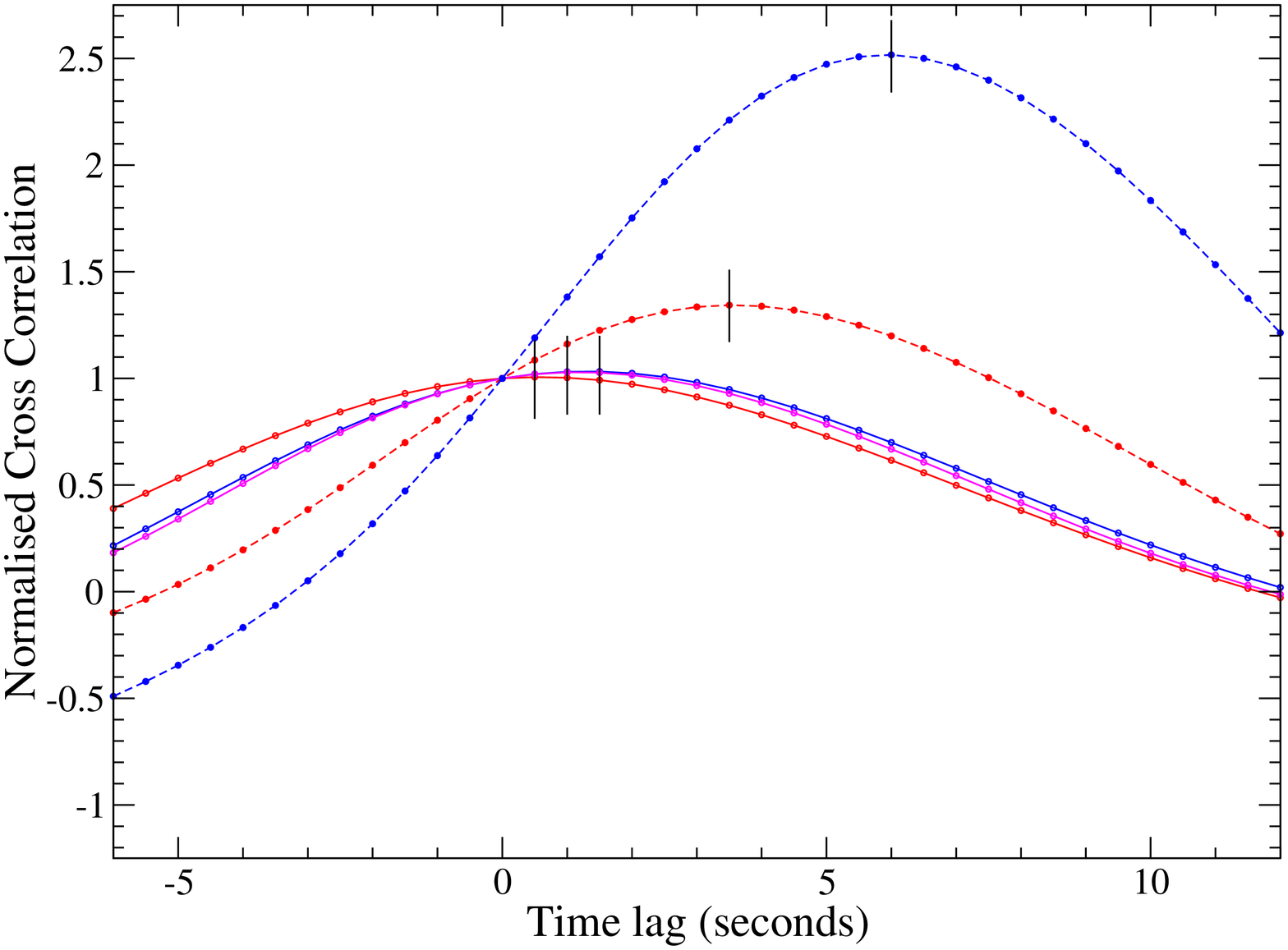}
\caption[]{{\it Left panel:}
Simulated light curves: the thick black curve represents the one at low energies; 
the others show three different examples of higher energy data. 
The  left-hand curves show the case when leading sides are delayed with respect to the 
low energy, but the trailing sides are the same in two cases and slightly different in 
the third one. 
In the right-hand curves, different shifts between the pulse maxima and decays are introduced.
{\it Right panel:} 
The cross-correlation functions for the simulated light curves
normalised to the values of zero lag.
Colours are the same as the corresponding simulated data.
}
\label{cross}
\end{figure}

A precise definition of this effect is not simple.
To  illustrate this, we consider some simulated, 
one low-energy,  and three high-energy, light curves as shown in Fig.~\ref{cross} (left panel). 
In the three examples on the left-hand side, high-energy curves have the leading sides delayed 
with respect to the low-energy curves, but the trailing sides are the same in two cases and slightly 
different in the third one. 
In the other two examples (on the right-hand side of the right panel in Fig.~\ref{cross}),
we considered different lags between the pulse maxima and the following decays.
It is apparent that delays are not constant during the bursts and reach the highest 
values at the transition between the $SLT$ and the $Pulse$.
Taking into account the conventional definition of a signal width, we can consider
the $Pulse$ height at half of the maximum and define the HXDs as the separation between these
values at different energies.
Thus, the horizontal segments plotted between the couples of pulses in Fig.~\ref{cross} (upper 
panel) give an estimate of these delays, which are equal to 1.75, 4.0, and 5.25 s for the left 
three bursts and to 5.75 and 10.25 s for the other two, while peak shifts are 2.5 and 5.0 s.
We must also distinguish between the delays occurring in the rising and decaying portions of the burst, 
hereafter indicated as HXD$_1$ and HXD$_2$, because they can be different.
The evaluation of lags between signals in two simultaneous time series is generally performed by 
means of their cross-correlation function (CCF). 
This method  is not sensitive to the variables we are working with and other algorithms must be applied.
This is clearly shown in the CCF plots of the above simulated data in the right panel of 
Fig.~\ref{cross}: CCF maxima for the first set of three burst data sets have very low values
in the range 0.5 -- 1.5 s, which are much smaller than the separation between pulses. 
In the other two examples, CCF maxima are clearly evident, but their lags are 3.5 and 
6.0 s, slightly higher than the corresponding peak shifts, but even much lower than the separations
at half peak heights.
These examples show that the CCF method does not always provide a proper evaluation of HXD: it is much
more sensitive to phase lags at the fundamental frequency, where it is the highest power, but it does
not measure well lags at higher harmonics, which are important when the $Pulse$ width is changing,
as in our case. 
In addition to CCF, we have therefore considered other methods which are
more appropriate for a self-consistent description of the burst inherent complexity.

\section{Limit cycle mapping in the {\it CR-E} plane}

\begin{figure}[h!]
\includegraphics[height=8.5cm,angle=-90,scale=1.0]{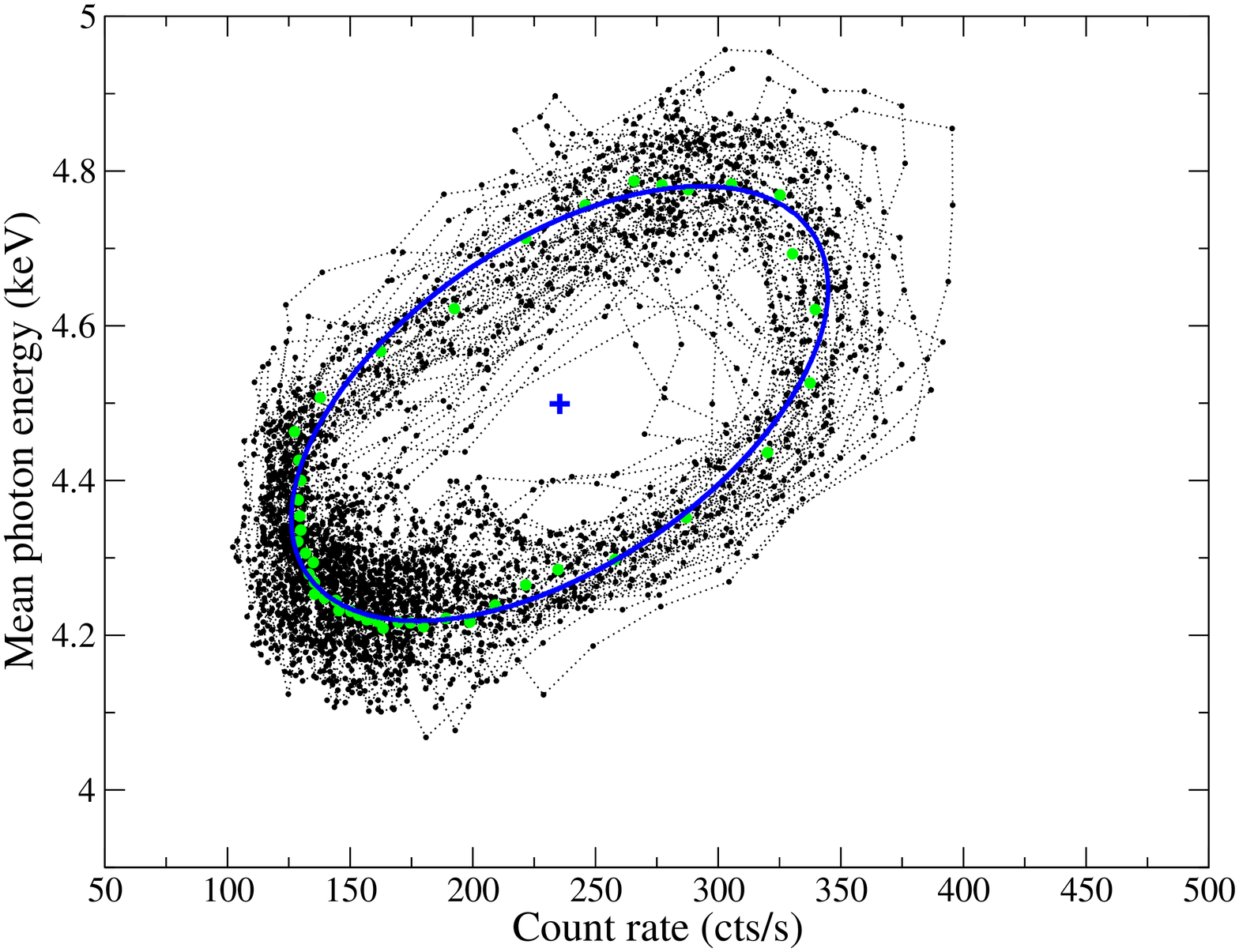}
\includegraphics[height=8.5cm,angle=-90,scale=1.0]{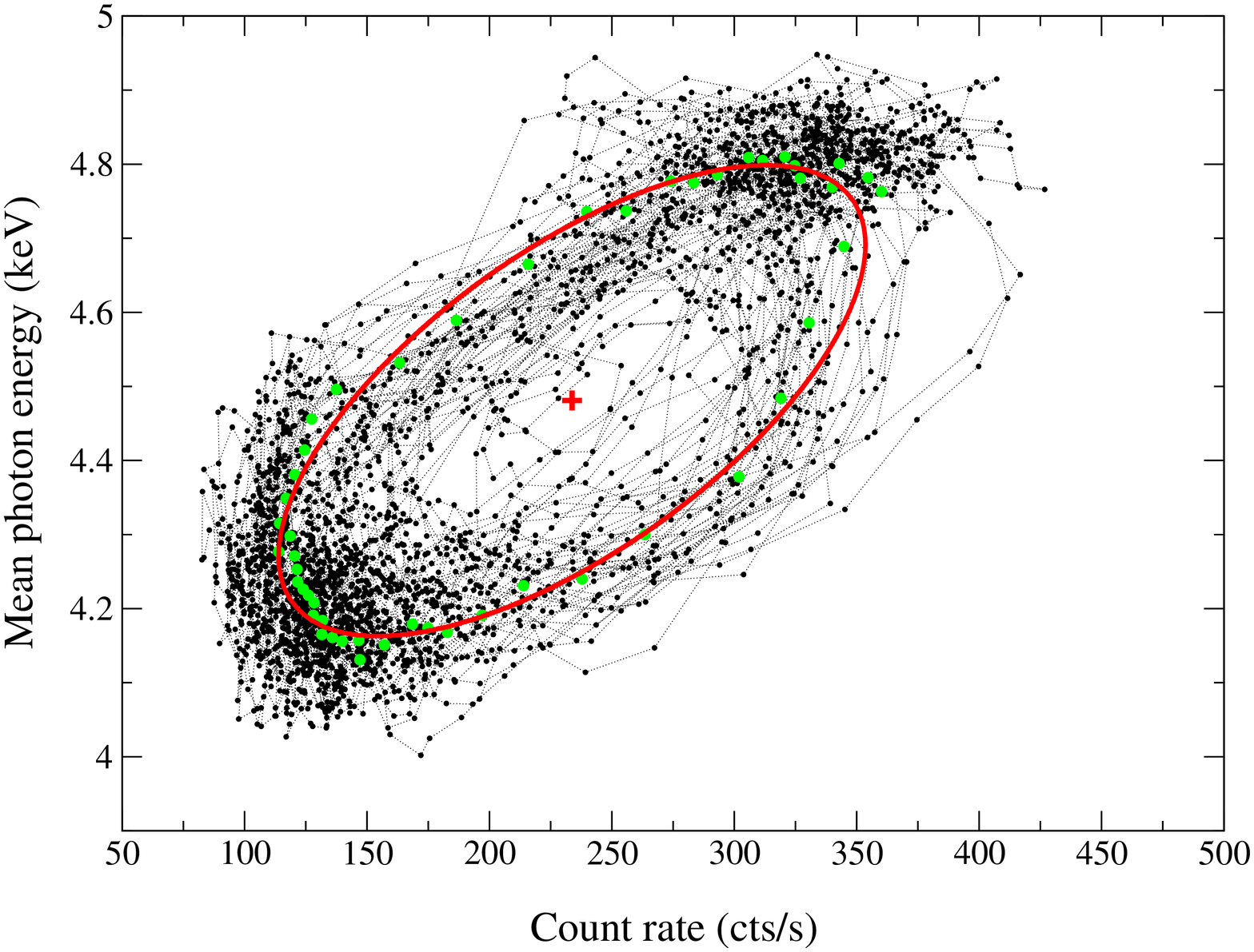}
\includegraphics[height=8.5cm,angle=-90,scale=1.0]{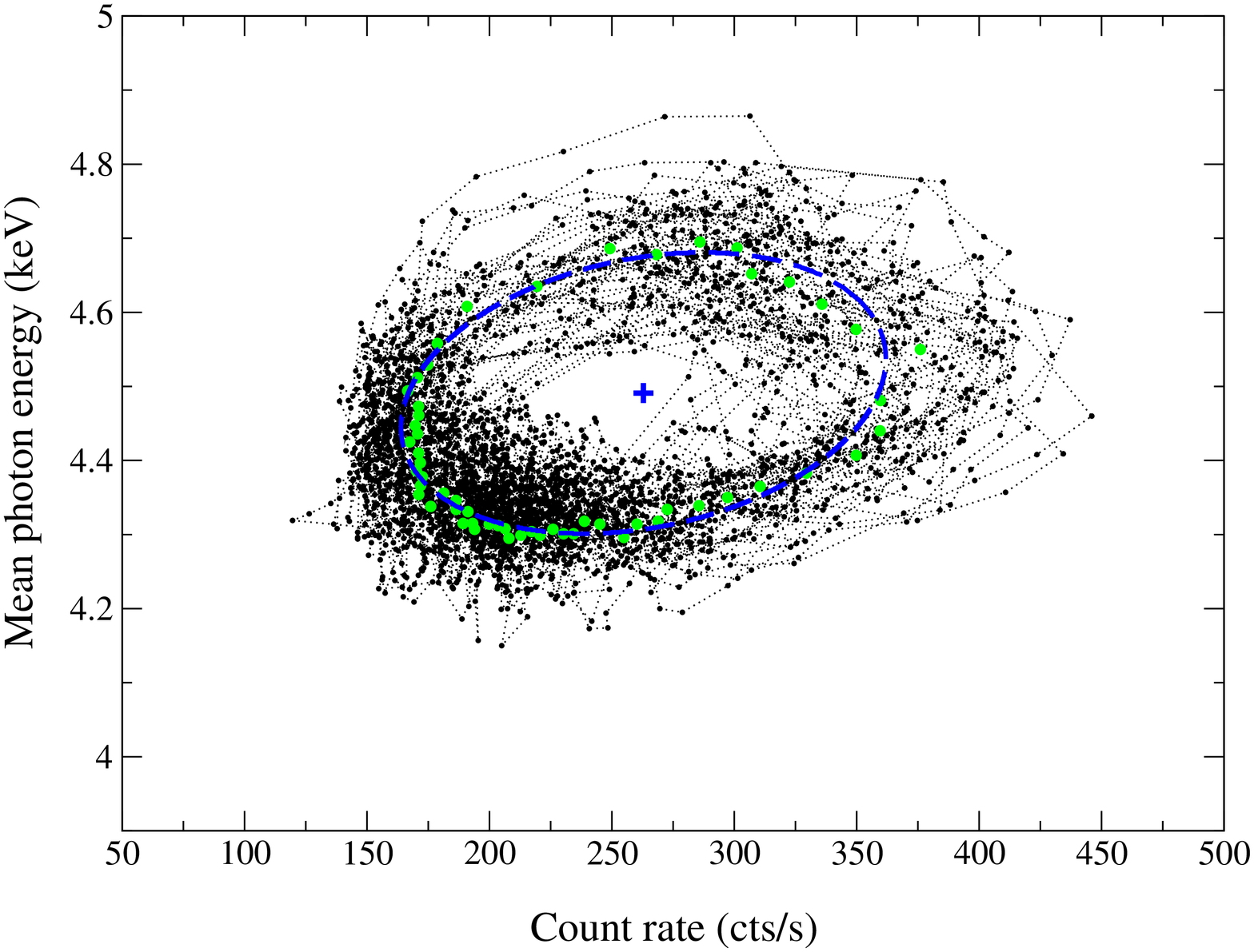}
\caption[]{
Trajectories for the data of the A8b (top panel), E5 (central panel),
and F7 (bottom panel) time series in the {\it CR-E} plane.
Dotted lines connect consecutive data points, the green dots are the mean
values in the set of angular sectors along the loop, and thick lines are
their best fit ellipses, computed as described in the text; crosses mark 
the centroids of the trajectories.
}
\label{fig3}
\end{figure}

To study the evolution of recurrent signals, it is useful to analyse the 
trajectories described in a suitable parameter space defined by two or 
more variables.
It is therefore necessary to have at least two simultaneous time series of 
independent quantities.
For dynamical systems, one could consider, for example, the time derivative of 
the original data series as an independent variable.
In our case, however, because of rather large fluctuations of the signal, 
the calculated derivatives present frequent sign changes and the resulting 
trajectories are highly confused.
We  thus preferred to use a dynamic space, having the count rate 
and the mean photon energy as coordinates (hereafter {\it CR-E} plane).
Hardness ratios (HR) could  be used equivalently instead of the mean energy, 
and we verified that the two variables are strongly correlated, as expected. 
We recall that the dominant emission component in the MECS range is the 
multi-temperature disk (see Paper II);  the mean photon energy can be related
to the temperature $T_{in}$ at the inner boundary of the disk, while the
count rate can be related to the integrated photon emission rate from the disk
(see Sect. 5.3).
For this reason we preferred to use the mean energy instead of HR. 
We note that these variables are statistically independent, at variance
with the plots of two hardness ratios  \citep[see, e.g.][]{vil_1, bel_2},
where there is an interdependence between the two
variables due to the use of count rates in a common energy band.
This parameter space cannot be considered equivalent to the phase
space used in the study of dynamical systems because the curves can 
intersect and can also be superimposed: different brightness states can
correspond to the same mean photon energy.
A more complete description of the physical state of the disk, in fact,
would require more than two variables.

For a better understanding of the evolution of a system in the {\it CR-E} plane,
one must consider that trajectories described by sources exhibiting variations 
with a one-to-one correspondence between the mean photon energy and the count 
rate are portions of a curve that becomes a straight segment when they are 
exactly proportional.
The occurrence of loops implies that there is a time shift between the two 
variables, and the direction of the motion along the loop depends on the sign of the
lag between the mean energy and the count rate.
This approach can be considered an evolution of the method that was originally proposed
by \citet{milne_1} for studying stellar variability, and some of the considerations
developed in this paper can be applied in the present analysis.

The {\it CR-E} trajectories for the A8b, E5, and F7 series are reported 
in the top, central, and bottom panel of Fig.~\ref{fig3}, respectively: the loop structure 
is clearly apparent in all the panels, but with some interesting differences.
First, we note that all loops are in a counterclockwise direction, which 
in our case corresponds to a lag of the mean energy series as expected from 
the occurrence of HXD, either in the {\it SLT} or in the decay of the {\it Pulse}, 
which is a stable characteristic of the $\rho$ class.
Data points are not uniformly distributed along the cycle: their density 
is generally much higher in the low-energy -- low-count rate part of the diagram.
This region corresponds to the {\it SLT} segment, while the {\it Pulse}, 
which completes the right-upper part of the loop, evolves on a shorter time-scale 
and has therefore a smaller number of points.
In the {\it CR-E} map of the E5 series, there is another high density region 
in the high-energy and high count rate part, corresponding to the high multiplicity
structure of the {\it Pulse}, and trajectories describe smaller secondary loops.
Our {\it CR-E} plots are strongly similar to that presented in Fig.14 of \citet{jan_2},
where they considered the two regions with a high density of points, which are also present in 
our maps. However, in that work the trajectories were not considered as a tool for the 
HXD evaluation.

\begin{figure}
\includegraphics[height=8.5cm,angle=-90,scale=1.0]{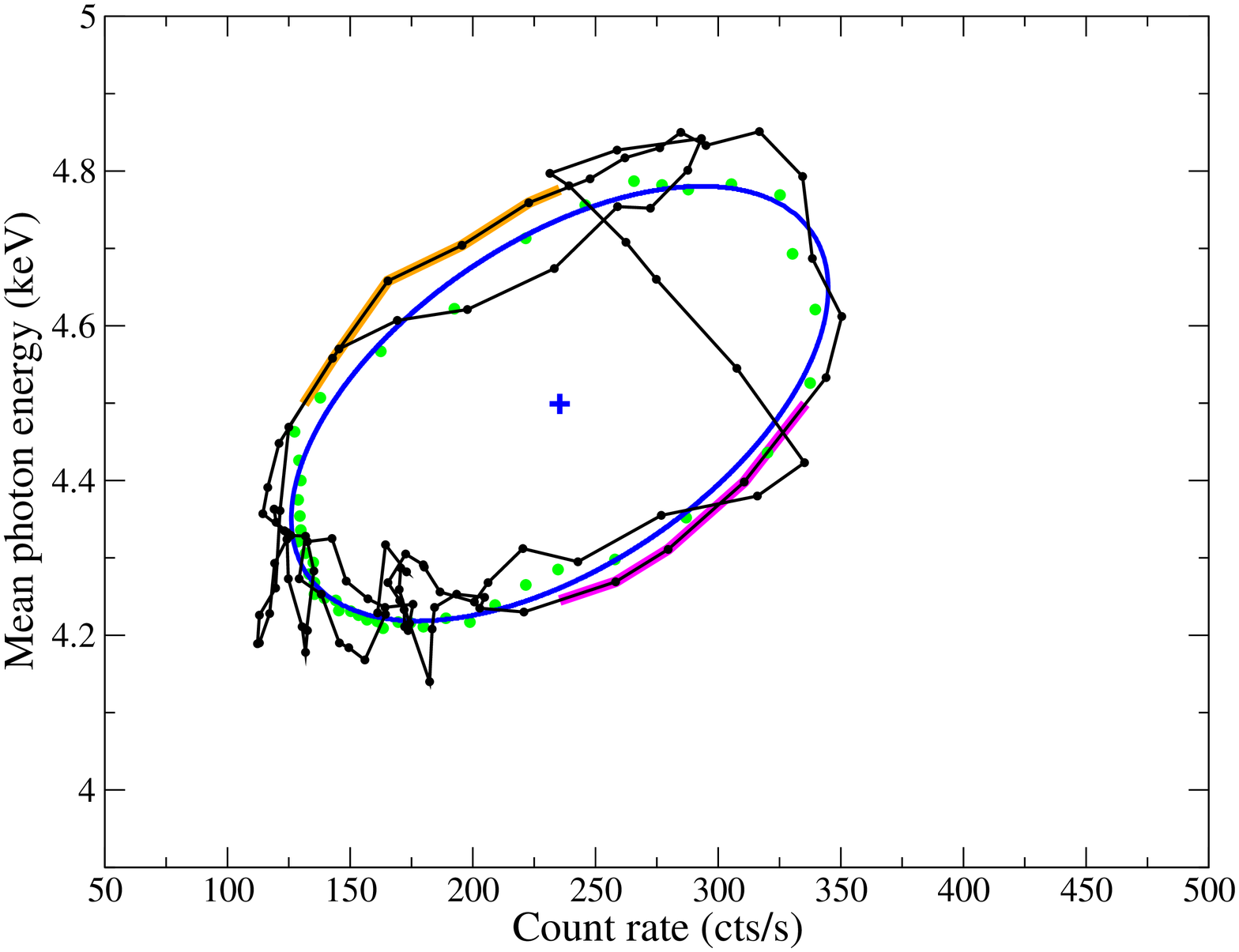}
\includegraphics[height=8.5cm,angle=-90,scale=1.0]{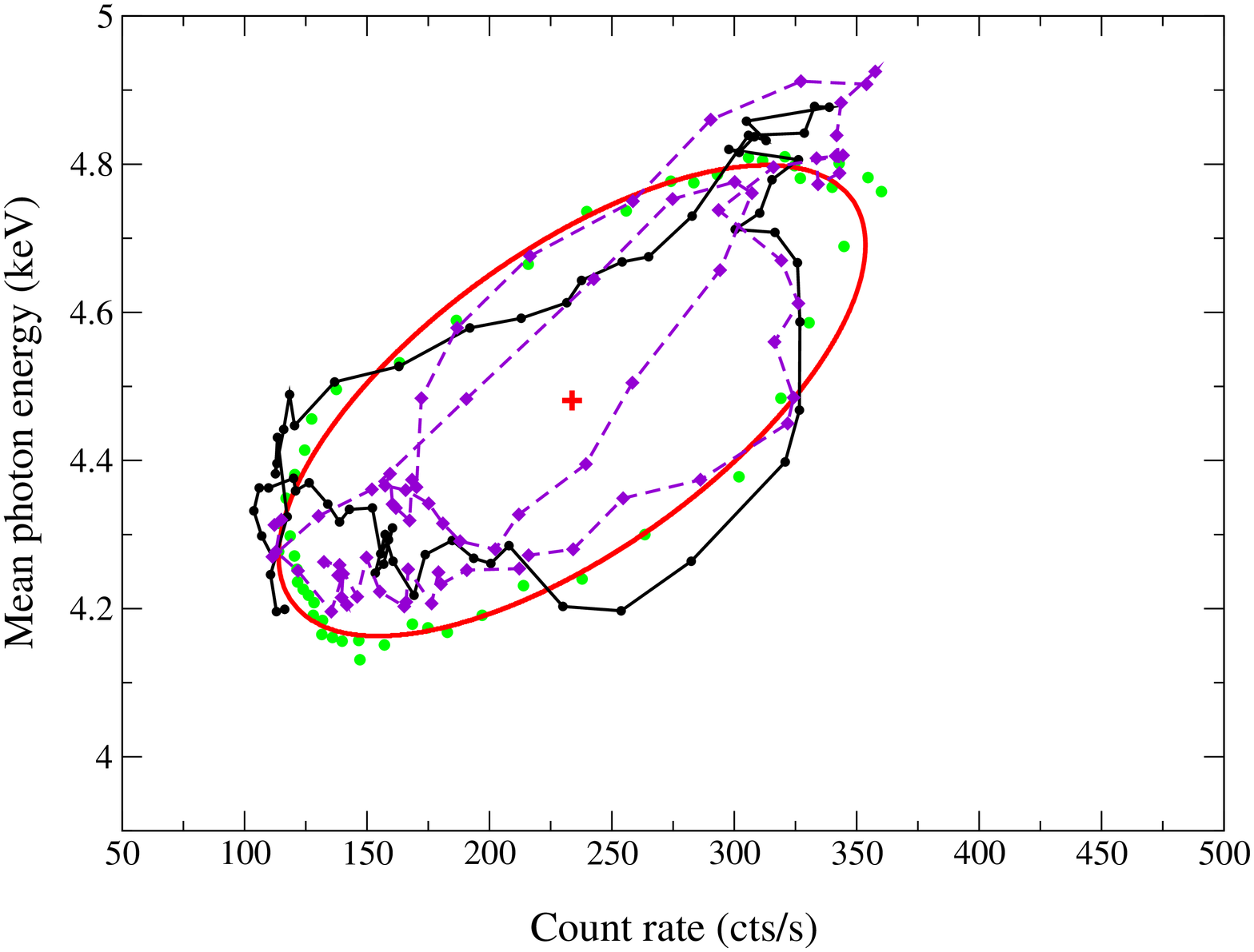}
\caption[]{
Trajectories in the {\it CR-E} plane of two subsequent individual bursts of the A8b 
(top panel) and E5 (bottom panel) time series.
Green filled circles and blue and red ellipses are the mean trajectories and the best
fits as in Fig.~\ref{fig3}; black lines connecting consecutive points track individual 
bursts, with the exception of the first ``anomalous'' burst in the bottom panel
plotted as a violet dashed line.
In the upper panel the thick black portions of the line connecting data points mark the
sections used to measure the HXD$_1$ (magenta) and HXD$_2$ (orange).
}
\label{fig4}
\end{figure}

\subsection{Mean trajectories: definition and main properties}

As a first step, we evaluated the central point encircled by the {\it CR-E} plane 
trajectories.
A direct evaluation of this point from a simple averaging of {\it CR-E} 
coordinates of all data points would produce a position that is close to the high-density 
region in the left-hand bottom side and not even encircled by several burst loops.
We therefore adopted the following procedure:
we assumed a preliminary guess for the point location and transformed the two 
variables by subtracting these approximate values and normalizing them by dividing the
resulting values by the respective standard deviations. 
From this point, we then drew a set of angular beams covering the entire plane,  
with an amplitude chosen to have in each of them a minimum number of 50 points;
we computed the mean of the radial distances of all points inside each bin and 
assigned this value to the central beam direction.
All these points track a first mean trajectory, and their mean values were used to 
calculate a new central point
(hereafter we  refer this point as the centroid of the loop; it is independent 
of the normalization and can be easily translated to the original {\it CR-E} plane).
This procedure was iterated until a convergence better than 0.5\% and a stable 
mean trajectory was reached,
The final trajectories are plotted as green dots in the panels of Figs.~\ref{fig3} 
and ~\ref{fig4}. 

All the mean trajectories present a nearly elliptical shape and so,
for each considered data series we computed the best-fitting ellipse; 
 the results were found to track them quite well, with some minor deviations. 
According to \citet{milne_1},  the elliptical shape can be understood on the basis
of the hexagon Pascal's theorem, assuming that there is a relation linking
the mean energy with the count rate.
Mean trajectories and best ellipses are also plotted for the {\it CR-E} maps in
Fig.~\ref{fig3} and their centroids are marked by a cross.
We note that the ellipses of the A8b and E5  series have similar orientation; 
the latter is only a bit more elongated along the major axis, whereas that of the F7 
series is smaller and has a different orientation with the major axis nearly aligned 
with the {\it CR} axis.
Moreover, from the plots of the  A8b and E5 series, which correspond to the regular 
and irregular mode, respectively, we see that their centroids remained very
stable (changes are smaller than 1\%). This indicates that different types of bursts 
do not significantly depend on these parameters. 
We also note  that, by our construction of the mean trajectories, they are
rather regular curves without minor loops, and very different from those originated
by peaks in bursts with high multiplicity. 
The information on structures related to the irregular mode does not appear in the mean 
trajectories, and their study must be based on the analysis of the trajectories 
of individual bursts (see Sect. 4.2).

\subsection{Trajectories of individual bursts}

As discussed above, the mean {\it CR-E} trajectories of all data series can be 
reasonably described by a regular curve like an ellipse.
Ellipses described with a constant angular velocity would result as a combination 
of simple harmonic motions on both coordinates, and the orientation ellipse would 
depend on their phase difference.
Time profiles of individual bursts, however, differ from a sinusoid on both variables 
and can exhibit several structures, depending on the {\it SLT} length and {\it Pulse} 
multiplicity.
Figure~\ref{fig4}  shows some examples of trajectories for a couple of bursts extracted from 
the regular  A8b series (upper panel) and the irregular E5 one (lower panel).
In the former case, we see a burst following  an elliptical loop rather well, while
the second one, with two well-separated peaks (multiplicity 2), has a trajectory 
that crosses the ellipse and presents a second clockwise small loop in the upper 
region.
As shown in Paper I, bursts in the regular series are characterised by a low 
mean multiplicity, which reaches 1.5 only in a few series.
Consequently, the region surrounding the centroids of the ellipse does not appear to be crossed 
by many trajectories.
Irregular series, like E5 (lower panel in Fig.~\ref{fig4}), which have a typical mean multiplicity 
close to or higher than 2.5, tend to fill this central region.
Moreover, they frequently present one or more secondary loops in the in the upper right-hand 
corner of their ellipses.
The connections between this structure and the burst type are also discussed  in \citet{neil_2}. 

\begin{figure}
\includegraphics[height=8.5cm,angle=-90,scale=1.0]{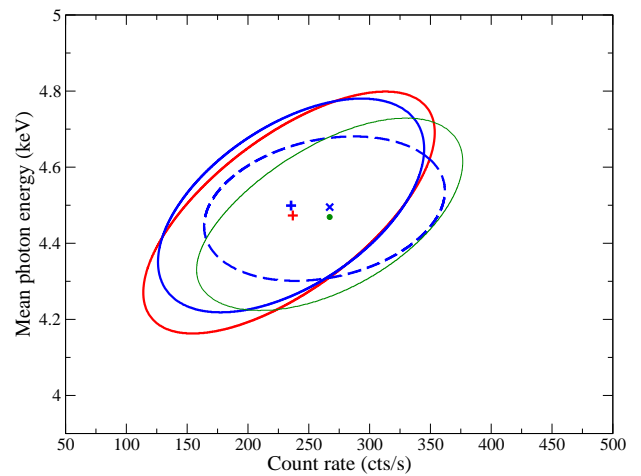}
\caption[]{
Best-fit ellipses of the mean trajectories in the {\it CR-E} plane of 
the A8b (blue solid line), E5 (red solid line), and F7 (blue dashed line) series, showing 
the change of the shape, orientation, and location of the centre of the ellipse of 
the last data series with respect to the previous two.
The green ellipse is computed from that of the A8b by increasing the count rate 
by the difference between the mean values with the F7 series and assuming that
the mean energy of this component was unchanged.
equal to 4.25 keV, as the typical {\it BL} 
value. 
}
\label{fig5}
\end{figure}

\subsection{Evolution of trajectories with the mean count rate}

The plots of Fig. 4 show that a change in the shape and orientation of the ellipse 
occurred with the transition between  interval II and III, when the 
the {\it BL} count rate increased (see Papers I and II).
This increase produced a shift of the centroid of the loop (and its mean ellipse) with 
respect to those of the preceeding series.
A change in the ellipses' orientation implies a variation of the phase difference of the
two variables. To make this effect clear, we investigated whether it can or cannot be explained
by a $BL$ increase.
To this aim,  we  modelled the burst structure as the sum of a constant {\it BL} level
plus a variable signal representing the rate variation along the burst. 
We indicate with $N_k=N_B+(N_V)_k$ the total numbers of photons in each time bin $k$,
where $N_B$ and $N_V$ are the  photons in the {\it BL} and variable component,
respectively. Assuming that the mean energy of {\it BL} photons remains stable at the  
value $\langle E_B \rangle$, the mean photon energy will be

\begin{equation}
\langle E_k \rangle = \frac {(N_B) \langle E_B \rangle + (N_V)_k \langle (E_V)_k \rangle}{N_k} 
\end{equation}

\noindent
where $\langle (E_V)_k \rangle$ is the mean energy of photons in the variable components 
in the $k$-th bin. 
In this simple model, we assumed that $\langle E_B \rangle$ remained stable and equal to 4.25 
keV during all the observation in agreement with the disk/coronal temperature in the $SLT$-1 phase,
as found in Paper II.
For a change of the {\it BL} count rate from $N_B$ to $N_B' = N_B + \Delta$ and consequently
from $N_k$ to $N_k' = N_k + \Delta$, as observed in the II and III time intervals, 
the mean photon energy would change to

\begin{equation}
\langle E'_k \rangle = \frac {N_k}{N_k +  \Delta} \langle E_k \rangle + \frac {\Delta}{N_k + \Delta} \langle E_B \rangle 
\end{equation}
\noindent
We note that this transformation is not a linear function in  $\Delta$, and therefore the 
ellipse will be changed in another closed curve, but such a modification would be relevant 
only for high $\Delta$ values.

We applied this transformation to the A8b ellipse for $\Delta$ equal to the change 
of the mean count rates from A8b to F7 series and plotted the resulting curve together
with the mean ellipses of the three  series considered in Fig.~\ref{fig5}:
its position was similar to that of F7, but the orientation results  changed slightly.
We conclude that an increase of the {\it BL} count rate cannot explain the 
change in the orientation of the ellipse observed in the two regular modes in the time 
intervals I and III and that a phase difference between the two variables must also occur.
Only for quite large $\Delta$ values is a clear change in the ellipse shape and orientation
obtained, but in this case a large change of the highest count rates is also found
in contrast with the observational results (see Fig.~\ref{fig3}).
This result also agrees  with the conclusion derived from the spectral analysis
from Paper II and \cite{neil_2}  that  different  modes require different disk 
temperatures and delays.

\section{HXD evaluation and results}

We adopted two different approaches to evaluate the HXD.
A first method, called the direct method, is based on the delay between the 
rising/decaying portions of individual bursts in the two series from their {\it CR-E} 
trajectories.
The second method is the calculation of the cross-correlation function between the count 
rate and mean energy time series.
This standard method was used as a benchmark for the results obtained with the limit cycle mapping.

\subsection{Direct method results}

\begin{figure}
\includegraphics[height=8.5cm,angle=-90,scale=1.0]{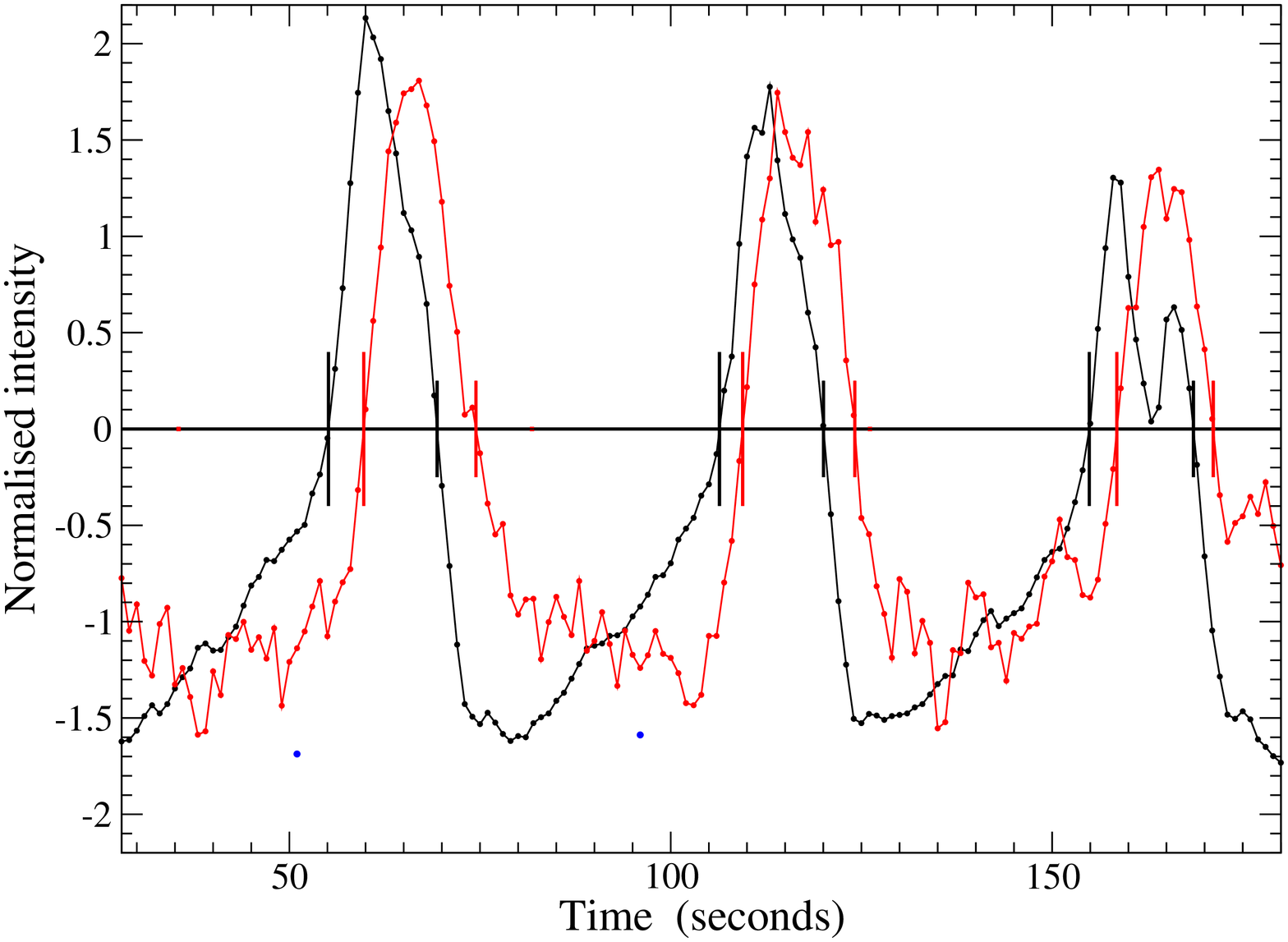}
\includegraphics[height=8.5cm,angle=-90,scale=1.0]{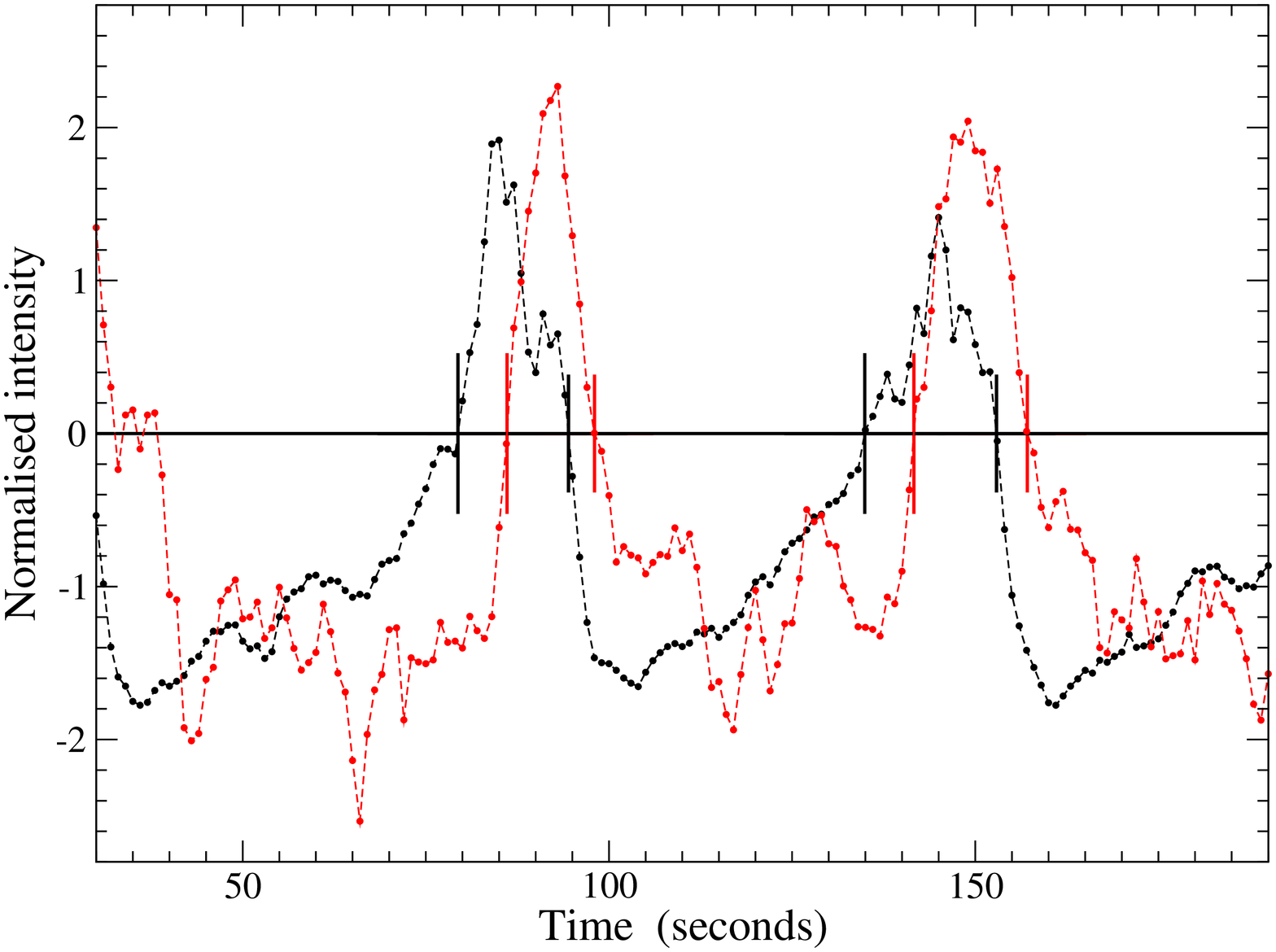}
\caption{
{\it Upper panel}:
A segment of the  A8b series illustrating how HXDs are evaluated in single bursts.
Black and red data are the count rate and mean energy series, respectively.
The zero level corresponds to the values of the centroid of the ellipse in the
$CR-E$ plot. 
Long and short vertical bars mark the time intervals of HXD$_1$ and HXD$_2$, respectively.
{\it Lower panel}: 
A segment of equal duration of the F7 series. We note that both HXDs are longer than
in the  A8b series.
}
\label{fig5bis}
\end{figure}

To estimate the  HXDs  from the {\it CR-E} trajectories, we subtracted 
the corresponding values of the loops' centroid from the count 
rate and mean energy data series and normalised the results by dividing  
the respective standard deviations.  
Then, we computed for every burst the time differences between the rising edges at their zero level 
to estimate HXD$_1$ and between the corresponding decaying edges to estimate HXD$_2$.
In practice, HXD$_1$ corresponds to the time necessary to cover the part of the
trajectory marked by a thick magenta line in the upper panel of Fig.~\ref{fig4}, 
and HXD$_2$ corresponds to the time needed to cover the orange line.
We representatively show a short segment of normalised count rate (black) and
mean energy (red) curves used for the evaluation of  HXD$_1$ and HXD$_2$ 
in Fig.~\ref{fig5bis}.
Both curves have a zero level corresponding to the centroid's values of the loop. 
The time difference between the two curves at zero is  HXD$_1$ in the $SLT$
and HXD$_2$ in the $FDT$. 
These two examples clearly show the presence of a time lag between
the maxima of the count rate and the mean energy series.
This lag is due to the fact that the energy distribution of photons in the $FDT$ 
has a temperature that, on average, is higher than the ones found in 
 $SLT$ and $P$-1; this effect lasts until the $BL$ level is reached 
(see Paper II).

\begin{figure}
\includegraphics[height=8.5cm,angle=-90,scale=1.0]{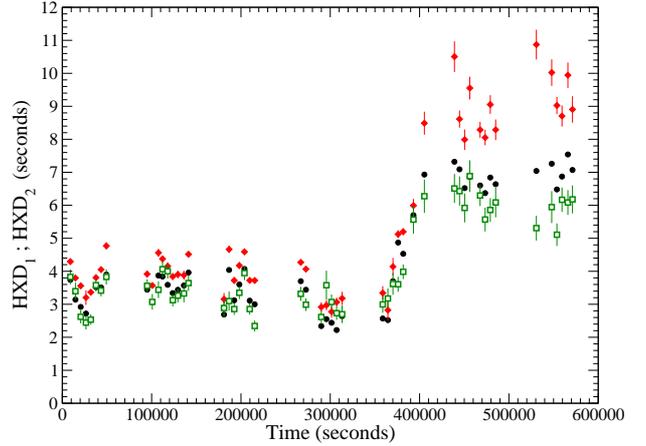}
\caption[]{
Evolution of the HXDs, measured by means of the various methods, between the
count rate and the mean energy data series during the entire time interval: 
results from the direct method are red diamonds (HXD$_1$) and green open squares
(HXD$_2$). 
Black filled circle are the HXD values obtained from the cross-correlation method.
The rather sharp increase from the interval II to III (360,000 to 400,000 seconds), 
after the {\it irregular} mode,corresponds to an increase in the mean  
count rate (see Paper I).}
\label{fig6}
\end{figure}

These values were then averaged over every time series; the resulting HXDs are
given in Table~\ref{table1} and plotted in Fig.~\ref{fig6} as red diamonds (HXD$_1$)
and open green squares (HXD$_2$).  
In the regular series, HXD$_2$ values are lower than the HXD$_1$ ones, with an average
difference of 0.6 s in the interval I that increases to 3 s in interval III.
Only in three irregular series did we find HXD$_2$ longer than HXD$_1$.
We note, however, that the estimate of HXD$_2$ is more uncertain than HXD$_1$ because of the 
irregularity of $Pulse$ decays due to the presence of several substructures, 
particularly in  bursts with high multiplicity.
When bursts exhibit such substructures, we assumed as HXD$_2$ the smallest 
time difference.

\begin{figure}
\includegraphics[height=8.5cm,angle=-90,scale=1.0]{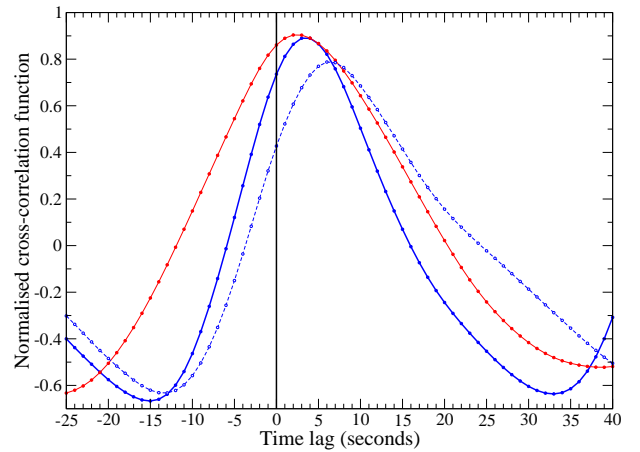}
\caption[]{
Cross-correlation functions of the count rate and mean energy data of three 
series, divided for their number of time bins: A8b (blue solid line), E5 (red 
solid line) and F7 (blue dashed line). 
We note a time lag varying from about 2.5 s to more than 6 s.
}
\label{fig7}
\end{figure}

\subsection{Cross-correlation results}

To compare our estimates of the HXD with the results reported in the literature,
we also computed the cross-correlation function (CCF) between the count rate and the
mean energy for all the  time series considered.
The CCF functions of the three sample series are plotted in Fig.~\ref{fig7}: all show a clear time lag 
of the mean energy series with respect to the count rate.
The position of the first maximum is considered as an estimate of the mean shift 
between the two time series giving the best matching between them, so we cannot obtain 
information on HXD$_1$ and HXD$_2$ separately.
We used a polynomial best fit to evaluate the position of CCF maxima, and the 
resulting values are given in Table 1 and  plotted in Fig.~\ref{fig6}.
The CCF results are lower than HXD$_2$ or  between the two HXD values obtained 
by means of the direct method.

\begin{figure}
\includegraphics[height=7.7cm,angle=-90,scale=1.0]{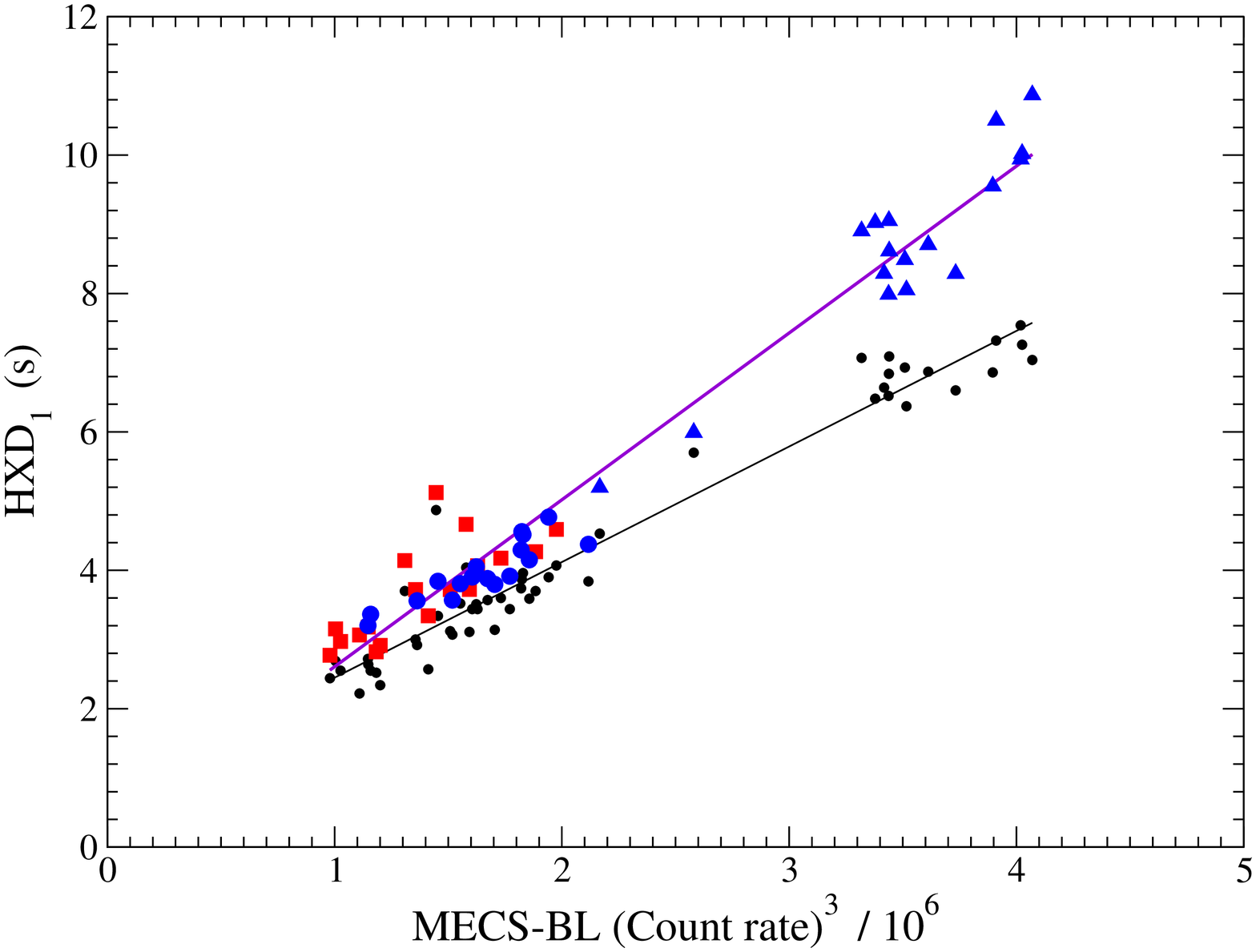}
\includegraphics[height=7.7cm,angle=-90,scale=1.0]{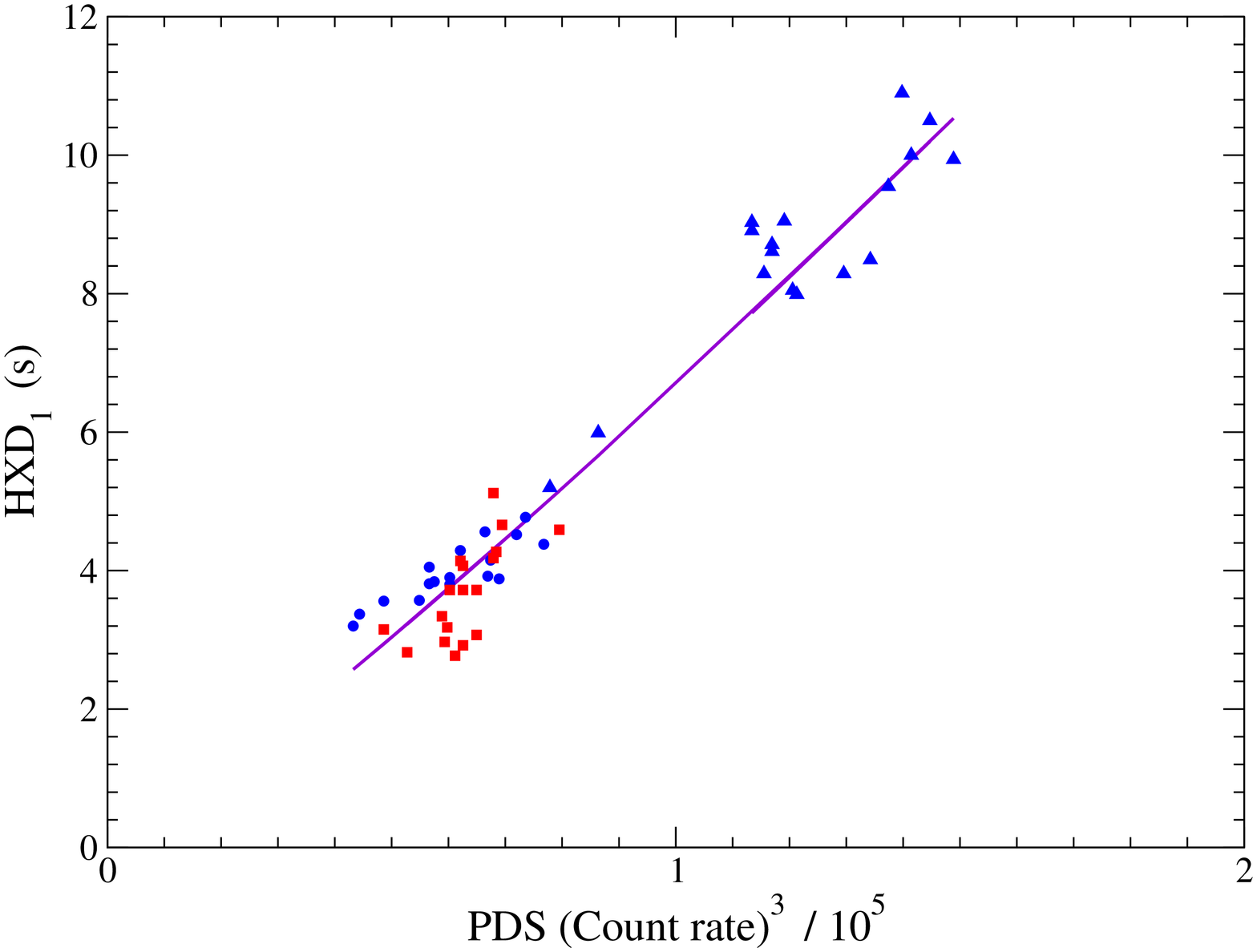}
\includegraphics[height=7.7cm,angle=-90,scale=1.0]{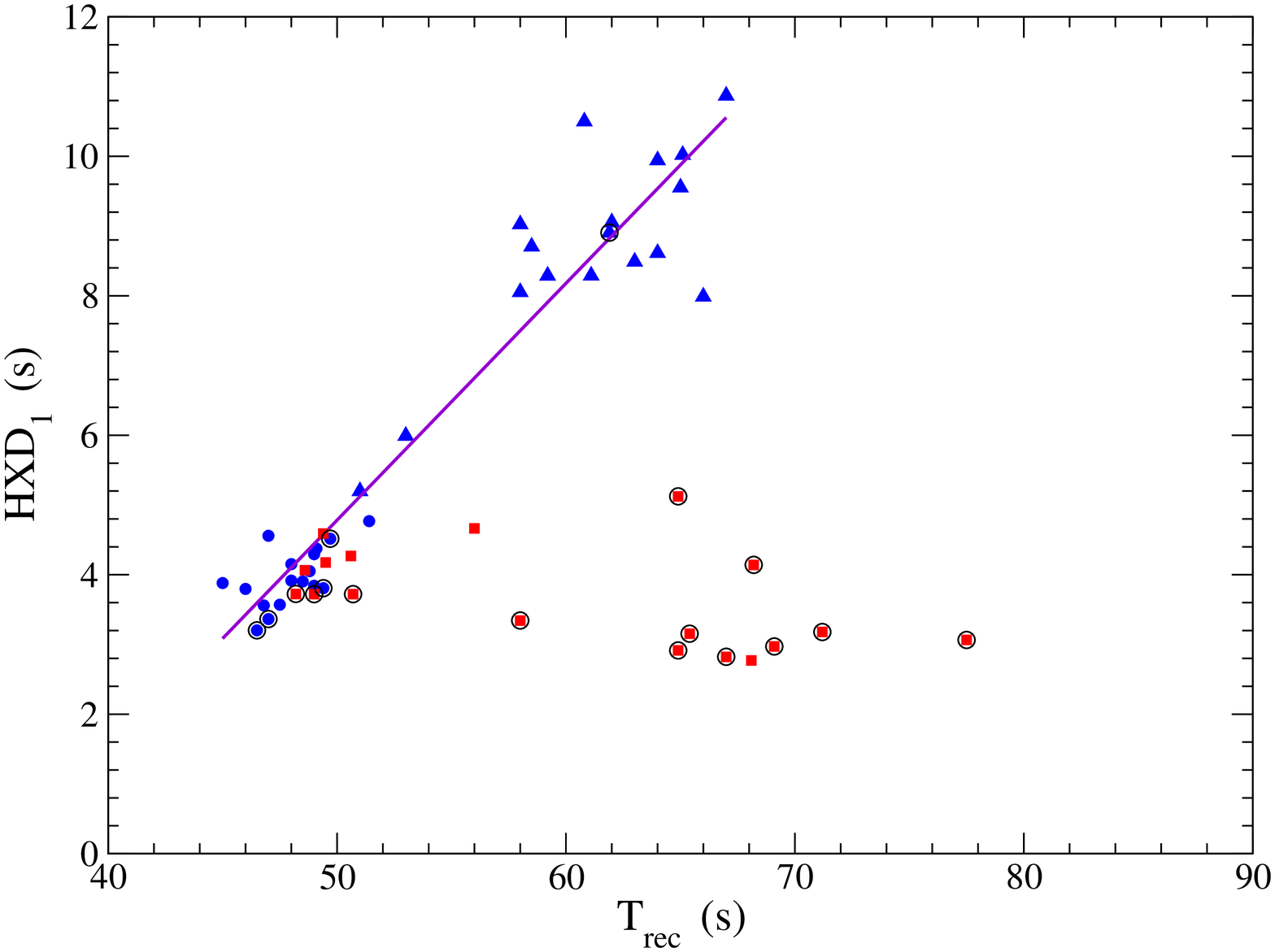}
\caption[]{
{\it Upper panel}: correlation between the HXD$_1$, evaluated using the direct method, 
with the third power of the mean count rate of the {\it BL} (in units of  10$^6$). 
Blue circles correspond to the interval I series, blue triangles to the  interval III series;
red filled squares correspond to the series in interval II.
The solid thick violet line indicates the linear best fit.
Black circles and the thin line are data points for the CCF estimates of HXD and their
{\it Central panel}: correlation between the mean HXDs with the mean count rate in the PDS band
(units of 10$^5$). Symbols are the same as in the upper panel.
{\it Lower panel}: correlation between the mean HXD$_1$ with the recurrence time of bursts as 
evaluated in Paper I. The solid thick line is the linear best fit of blue points only.
Black circled points mark the irregular data series.}
\label{fig8}
\end{figure}

\subsection{Evolution of the HXD}

The time evolution of the HXD measures is remarkably similar to that of the mean count rate and 
a very interesting behaviour emerges: 
the HXD remained practically stable within the rather narrow range [3--4] s during the 
interval I, when \grs~ was in the regular mode. A small decrease occurred in the interval II, 
during the irregular mode, which was followed by a rather sharp increase between 360 ks and 
400 ks to a mean lag value around 7 s, when the source turned again to the regular mode 
interval III.
As shown in Fig.~\ref{fig6}, this evolution of HXD in the course of the observation is 
remarkably similar to those of other quantities that characterise the $\rho$ class (see Paper I),
 particularly  the {\it BL} count rate (see Fig. 15 in Paper I).
To show this, we use a double-log plot of HXD vs. count rate. 
A power-law fit with an exponent close to three  provides a good description of this apparent trend.
 This result is shown in the upper panel of Fig.~\ref{fig8}, where we plotted the 
values of the HXD$_1$, against the third power of the count rate (in units of 10$^6$ counts$^3$). 
The resulting linear correlation coefficient is equal to 0.982. 
We note that this exponent value depends on the method used to estimate the HXD;
in fact, by repeating the same analysis with the values derived from the CCF method, we obtained 
a power law index closer to 2.5. 

In the upper panel of Fig.~\ref{fig8}, these values are also plotted together with the
linear best fit, corresponding to a linear correlation coefficient equal to 0.974.
According to the timing analysis of Paper I, this result implies that HXD$_1$
is also related to the mean count rate of PDS (15--100 keV) and to the recurrence time of  
the bursts.
The former correlation is shown in the central panel of Fig.~\ref{fig8}, where  HXD$_1$  
is plotted against the third power of the PDS count rate (the linear correlation coefficient is 0.973), 
while the one with the recurrence time is shown in the lower panel of the same figure.
In this case, however, it is important to distinguish regular series from the irregular ones: 
in the latter mode, in fact, it is very difficult to evaluate a reliable value 
of $T_{rec}$.  
Fourier periodograms of irregular series are generally of M type (Paper I) with several 
prominent peaks, and the central value of the period range in which 
they are present was used in the plot.
These series, which occurred mainly in interval II, appear to be characterised 
by rather short HXDs. 
For the series of intervals I and III, which are mostly regular, we again obtained  a very high 
linear correlation coefficient equal to 0.958; the corresponding best fit is plotted in 
Fig.~\ref{fig8}. 

\begin{figure}
\includegraphics[height=8.5cm,angle=-90,scale=1.0]{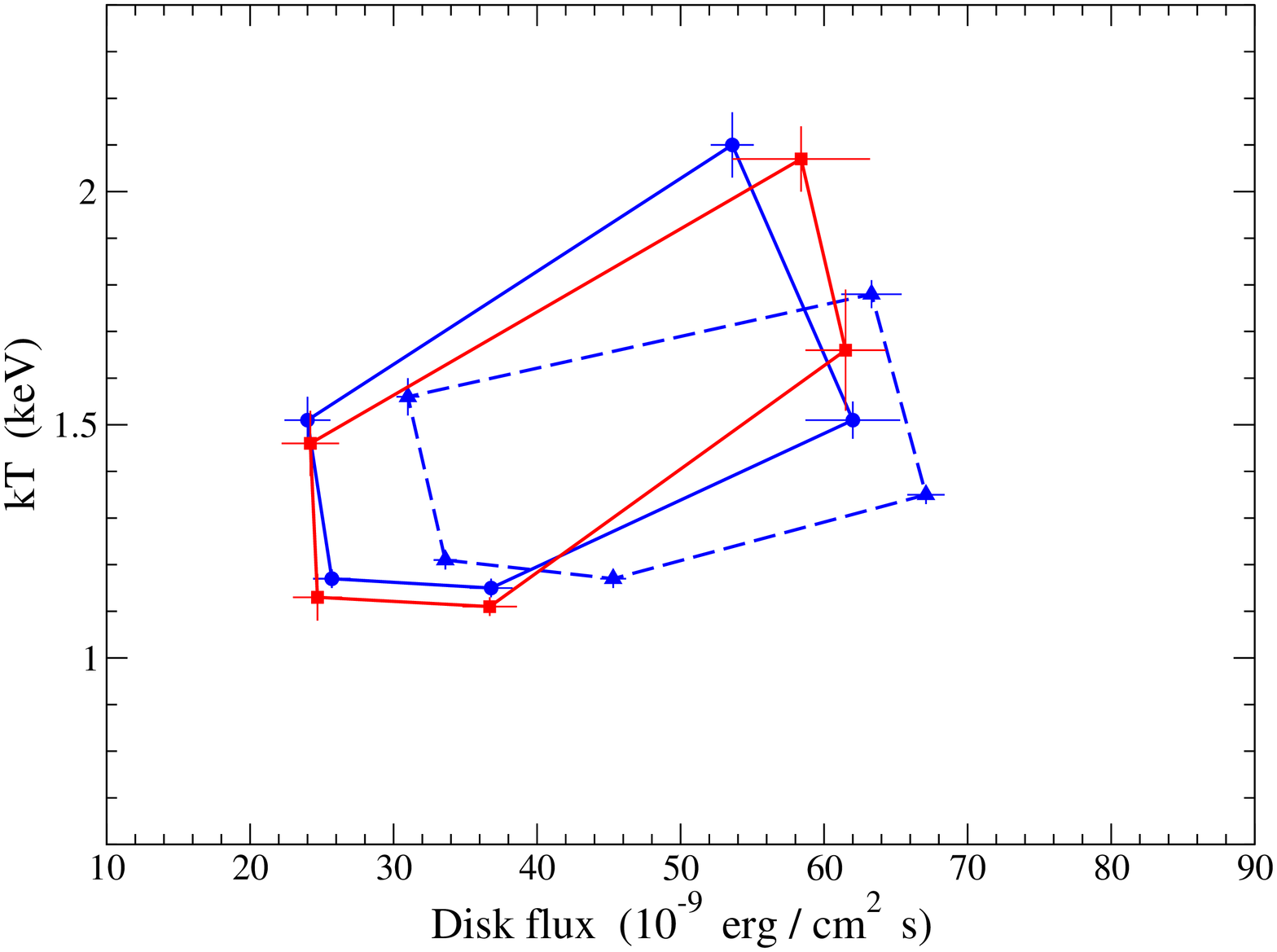}
\includegraphics[height=8.5cm,angle=-90,scale=1.0]{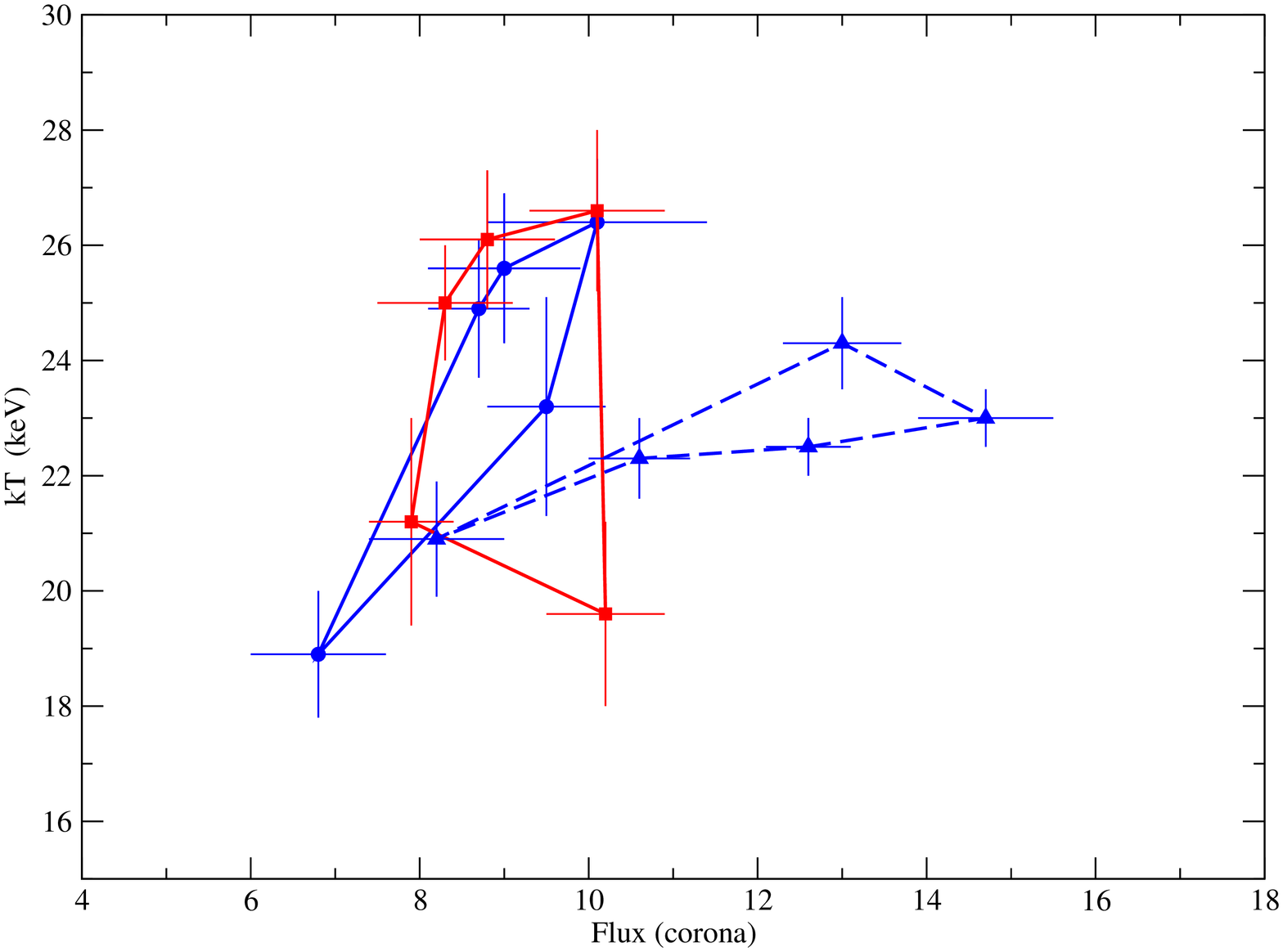}
\caption[]{
Upper panel: Evolution of spectral parameters of the disk component (see Paper I, 
Table 2) in the five segments of the bursts in the three intervals: I (solid blue 
line), II (solid red line), III (dashed blue line). \\ 
Lower panel: Evolution of spectral parameters of the coronal component (see Paper I, 
Table 2) in the same five segments of the bursts as in the upper panel.
}
\label{fig9}
\end{figure}

Loops in the {\it CR-E} plane can also be depicted in terms of corresponding spectral
parameters \citep[see also][]{neil_1,neil_2}. 
In Paper II we evaluated temperatures and fluxes for the multi-temperature disk and
a surrounding hot corona component in five burst segments; their plots are given in 
Fig.~\ref{fig9}.
The general structure of the loops in the {\it CR-E} plane of Fig.~\ref{fig3} and 
\ref{fig9} can be easily recognized in the disk plot. 

\begin{figure*}
\center{
\includegraphics[height=7.5cm,angle=0,scale=1.0]{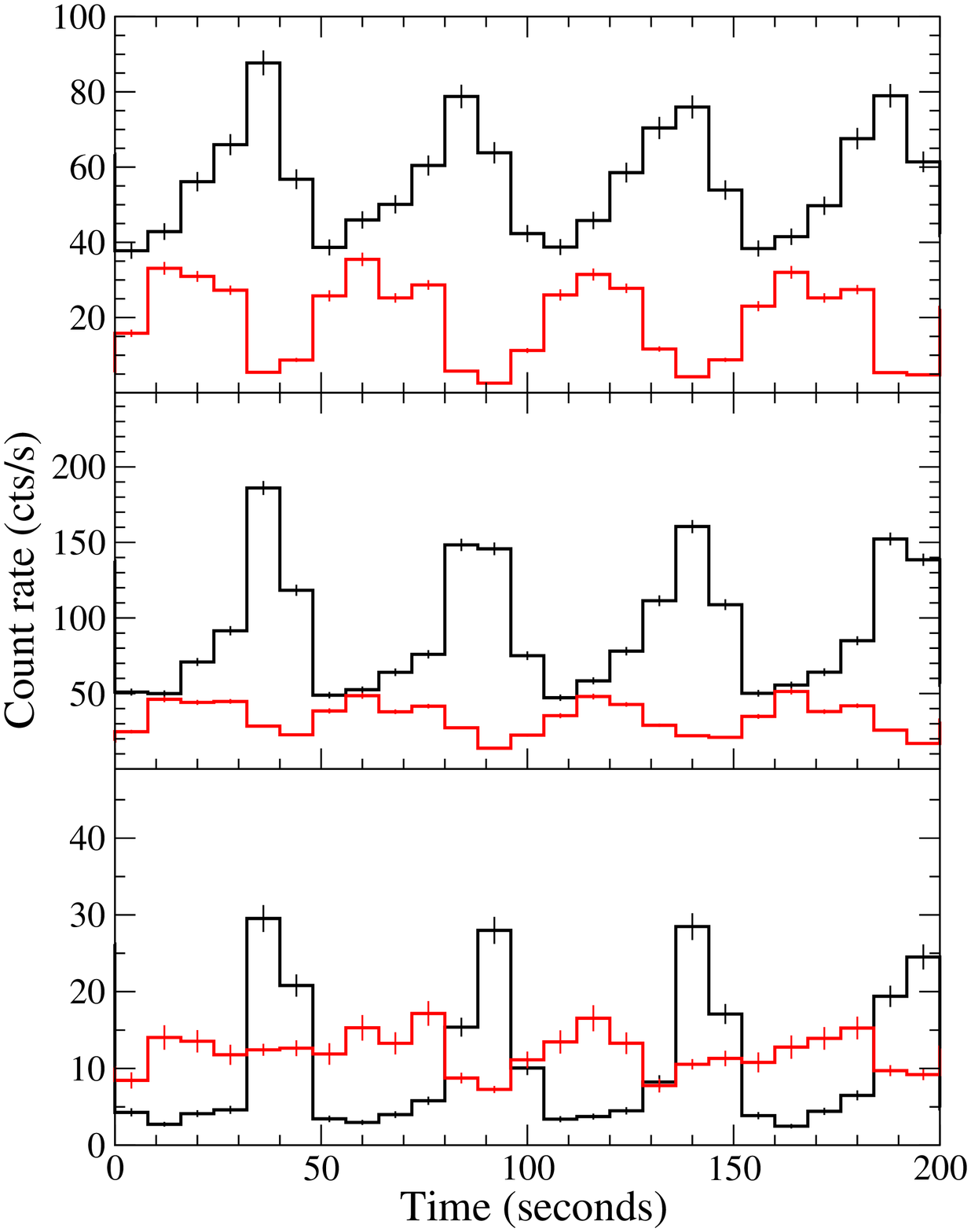}
\includegraphics[height=7.5cm,angle=0,scale=1.0]{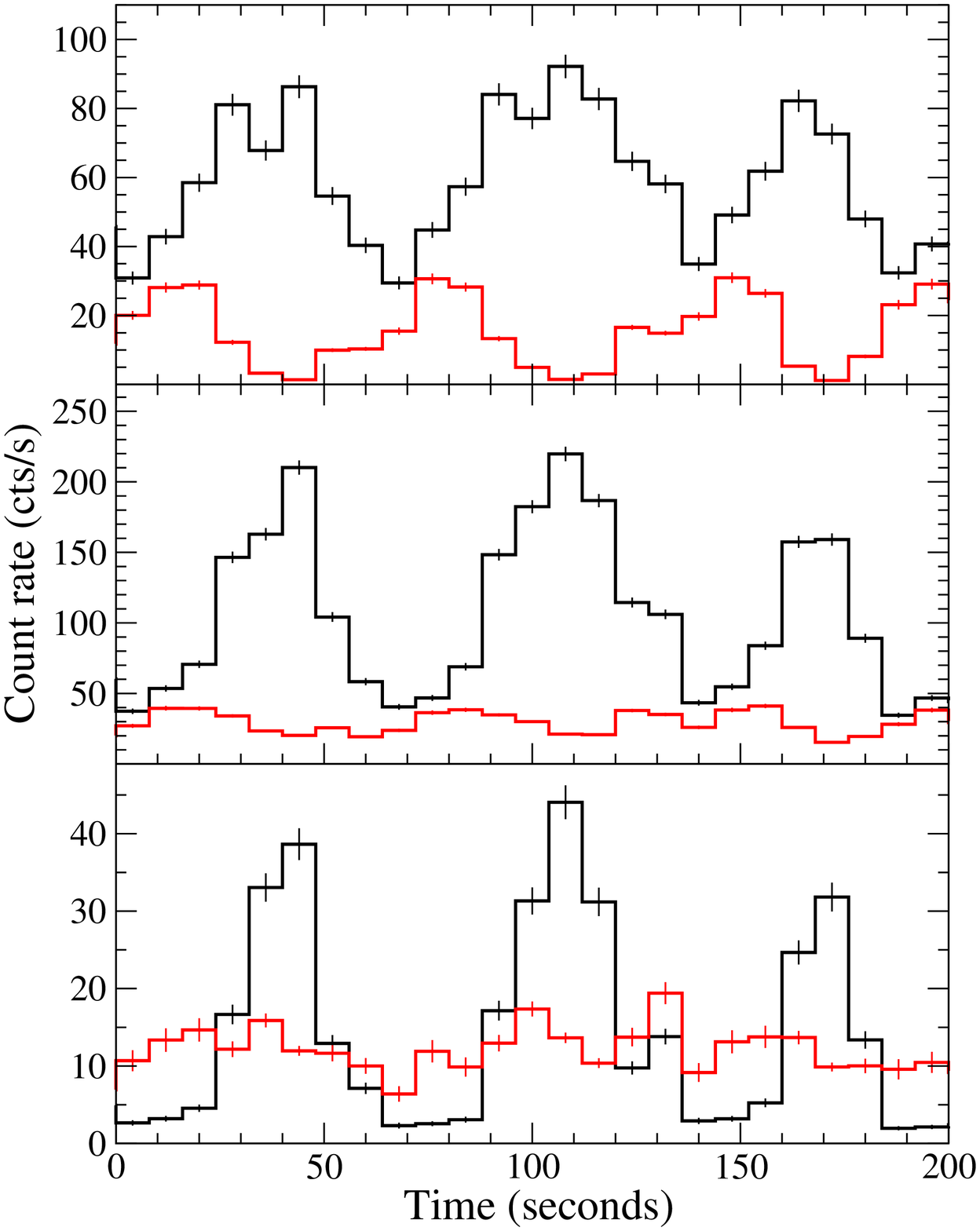}
\includegraphics[height=7.5cm,angle=0,scale=1.0]{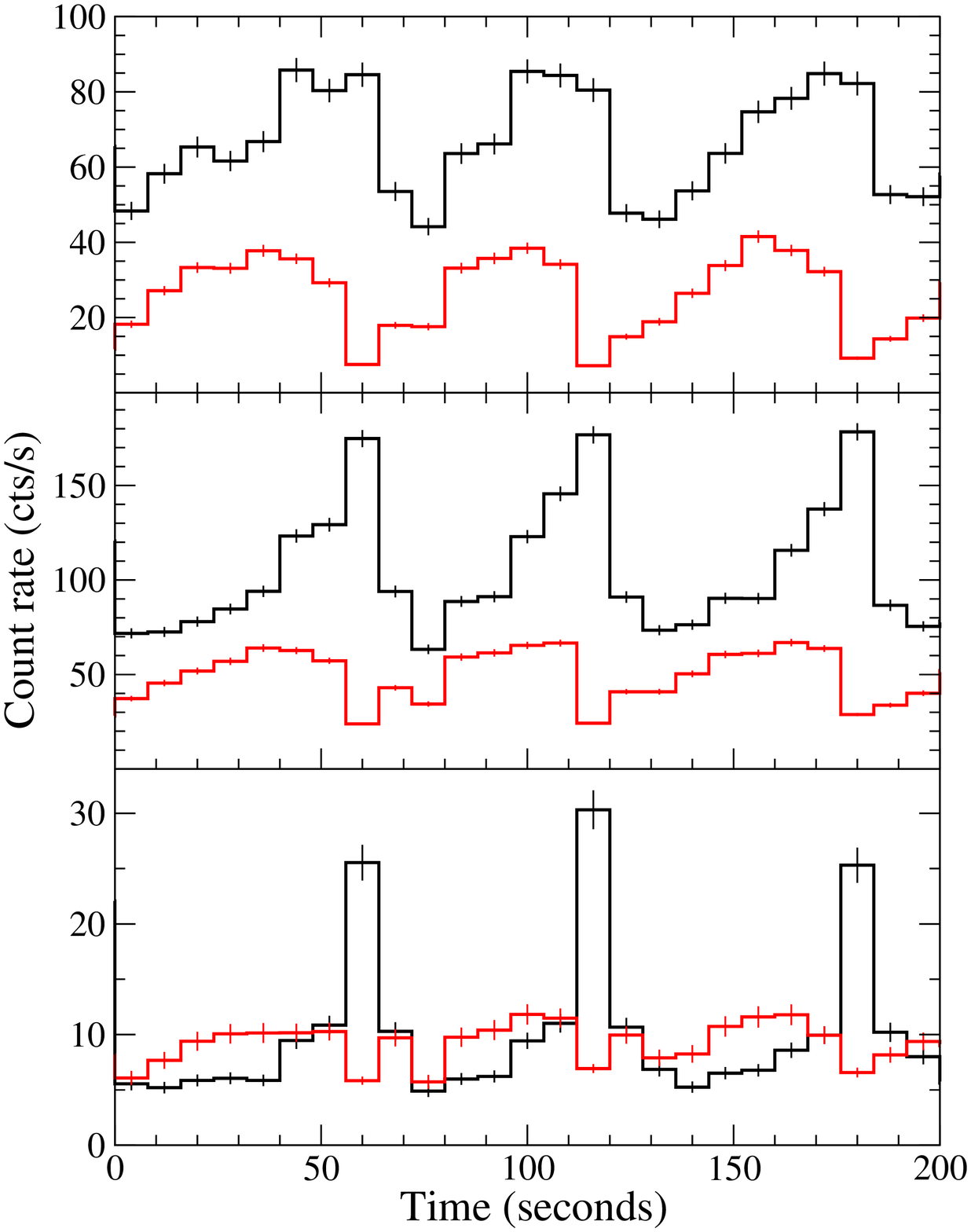}
}
\caption[]{
Short segments 200 s long of the count rates of the light curves for the multi-temperature 
disk (black) and the corona (red) emission components in the three energy ranges [1.7--3.4]
(top), [3.4--6.8] (center), and [6.9--10.2] keV (bottom) for the three series A8b (left panel),
E5 (central panel), F7 (right panel). Rates relative to the corona were multiplied by 8
in the  energy range [1.7--3.4] keV and by 2 in the other ranges to superimpose the light curves on
the disk ones.
}
\label{fig10}
\end{figure*}

Using the results of the spectral analysis presented in Paper II, we constructed 
the light curves for the multi-temperature disk and for the corona emission.
They were obtained from the detected rates according to the following formula
\begin{equation}
R_{comp}(t)  =  R_{total}(t) \, \frac {f_{comp}}{f_{total}},
\end{equation}
\noindent
where $R_{comp}$ is the rate relative to each spectral component at the time $t$, $R_{total}$
the detected rate, ${f_{comp}}$ and ${f_{total}}$ are the component and total fluxes
in the considered energy range. 
These factors depend on the source spectrum and change along the five burst segments that 
we identified in each series.
Short portions of the light curves for the multi-temperature disk and the corona for 
the three series A8b, E5, and F7, previously considered in this paper, are shown in the 
three panels of Fig.~\ref{fig10}.
In each panel we plotted the computed count rates in three energy ranges [1.7--3.4], 
[3.4--6.8] and [6.9--10.2] keV.
We note that the {\it Pulse} appears only in the disk component curve (black curve) and 
that the high-energy peaks lag behind the ones in the soft range with a behaviour coherent with 
the  HXDs detected in our analysis.
The corona emission is rather stable, particularly at energies above $\sim$7 keV, while
in the lowest range it is strongly anti-correlated with the disk component.

\section{Discussion}

We have shown that trajectories in the {\it CR-E} plane provide a useful tool
for investigating the complex behaviour of binary X-ray sources, particularly 
when they exhibit quasi-periodic variations as \grss does in the $\rho$ class.
An interesting parameter for the physical modelling is the delay of the X-ray
emission in the hard X-ray range with respect to the soft band.
We developed a method based on the study of these trajectories to measure the
HXD values observed in the burst sequence. 
Our results offer a wider perspective on the HXD phenomenon in \grss. In particular, 
we have shown that in the $\rho$ class the hard time lag can vary significantly, 
from $\sim$3 s up to $\sim$10 s, in tight correlation with the {\it BL} count rate 
and likely proportional to its third power.
This general evolution, despite some minor systematic deviations, is  consistent 
with the results obtained by means of the CCF between the count rate and the mean energy 
data series, which is sensitive to the phase lags between the two signals.

\citet{neil_2} reported in some RXTE observations a relation between HXD and burst 
multiplicity that is not  evident from the BeppoSAX data examined in this work.
This follows from the different definitions and methods used to evaluate the HXD,
because we focused our analysis mainly on HXD$_1$, which occurs in the transition between
the $SLT$ and the $P-1$ and is not expected to be related to the peak multiplicity.
Particularly interesting is the finding that the HXD$_1$ is
highly correlated with some important source parameters, such as the baseline level of the
stable component onto which bursts are superposed and the mean PDS count rate 
(see Table~\ref{table1}).

This result can imply that the development of the limit cycle, which is typical of $\rho$
bursts, is physically related to the state of the disk and that it can depend on 
some relevant quantities, such as the accretion mass rate and the local
temperature value.   
 
A phenomenon similar to HXD, as considered by us, is also observed at much lower 
frequencies in other disk accretion systems such as dwarf novae.
In the outbursts of some nova-like sources, it was noticed that the rise of high-frequency 
radiation follows that of the optical with a variable delay, ranging from 
minutes to days \citetext{UV:~e.g.~ \citealp{has_1,has_2,can_1};~ soft X:~ \citealp{WHB_1}}, 
while the decline in the two bands occurs almost simultaneously.

\subsection{Physical origin of the HXD}

The geometry and physical characteristics of the corona of \grss are still debated. 
One possibility is to set the corona at the base of a jet \citep{nob_2}, anchoring a 
source of variability in the accretion disk (like a flaring magnetic episode or 
a density wave) and adjusting in a suitable way the distance from the jet and its 
opening angle. However, such scenarios seems unlikely for the $\rho$ class bacause
it has been associated with 
unstable jet formation and generally weak radio emission \citep{kw_1}. 
A more plausible scenario could be connected to a hot, geometrically thick compact
corona and a truncated accretion disk, whose inner radius loops between
a maximum and a minimum (possibly associated with the last stable orbit radius) value. 
These models have been successful in reproducing the general light curve shapes typical of 
the $\rho$ class \citep{hon_1, szusmi_1, taa_1, jan_1} and  rely on the classical 
thermal instability model of \citet{ligear_1} as the agent of the cyclic behaviour. 
Another scenario, which is more physically motivated, is based on a geometrically thin 
and cold accretion disk (but with a variable and opportune prescription for the viscosity) 
and a corona above it \citep{nayakshin2000}. 

A first attempt to derive from this model an explanation for the hard lag phenomenon was 
provided by \citet{jan_2}, which presented a simple interaction model between disk and corona 
that provided, at least qualitatively, the observed time-scales. 
The model assumed a mass exchange between disk and corona, but the coronal thermodynamic 
equilibrium was simply set by the virial temperature (Compton processes are thus neglected). 
Besides providing the observed time-scale of the lag ($\sim$ 1 s), the model produced 
anti-correlated disk and coronal light curves, and predicted luminosity variations of the 
corona within a factor of two between the minimum and maximum fluxes. 
These predictions are in good agreement  with what we observationally found. 
The new observational facts that add now up are the correlation between the baseline count 
rates and the hard time lag and the wide spread in values of the observed lags. 

Finally, always within the instability model, it is possible to interpret the hard lag 
as a typical thermal time-scale of the region of the disk, which experiences the 
instability producing the limit cycle behaviour. 
While the viscous disk time-scale sets the recurrence time of the outbursts \citep{bel_1},
it is expected that the thermodynamic time-scale should be at least a factor 10 less, of the 
same order of the rising outburst time-scales \citep{nayakshin2000}. 
As shown in Fig.\ref{fig10}, the most variable component in the MECS energy range is 
always associated with the thermal disk emission.
We argue therefore that such hard lags could be mostly due to the thermal adjustment of the 
rising/decaying temperature within the inner parts of the disk, which are probably due to mass-accretion 
variations  propagating in the disk.

The regularity and the rather stable HXD, which are often observed in the $\rho$ class bursts, suggest
that it can be described by a set of differential equations, likely including one or more 
non-linear terms that combine the evolution of luminosity and temperature.
The parameters in these equations, in turn, could vary on time-scales different from that of the 
bursts, e.g. the mass accretion rate; they could be responsible for systematic changes exhibited by
the source and also produce changes of variability class.
A first attempt to construct a mathematical model for the $\rho$ class appears very promising
and, if confirmed,will be the subject of a forthcoming paper.

The discovery of another source, IGR J17091-3624, which also shows the heartbeat phenomenon
typical of the $\rho$ class of GRS 1915+105, opened the possibility to place much stronger 
constraints on the physical interpretation of such complex behaviours. 
This new source has shown so far several of the variability classes originally discovered 
in GRS 1915+105, albeit on faster time-scales and seemingly at lower accretion luminosities. 
Further and more detailed comparative studies of the two sources are needed to improve our 
understanding of the Eddington/radiation pressure instability as a function of disk (corona)
viscosity, accretion rate, optical depth, radiative efficiency, and mass transfer prescriptions. 


\begin{table*}[h]
\caption{Observation runs of the {\it Beppo}SAX pointing of \grss in October 2000 used in the 
HXD analysis.
The columns list observation codes, names of the runs, the time in seconds from the starting time of the
first run (20 October 2000 at UT = 21$^h$ 26$^m$ 55$^s$), the exposure time, MECS and PDS count rates 
in the 1.3--10 keV and 15--100 keV energy ranges, HXDs measured from cross-correlation analysis, the delay
measured in the rising portion of the burst HXD$_1$ with the Direct method,
the delay measured in the decaying portion of the burst HXD$_2$ with the Direct method, the number of bursts
the recurrence time (data in square brackets, for M type series, are the centres of the interval where 
several peaks appear in the Fourier periodogram - Paper I), and the type of data series.}
\label{table1}
\begin{center}
\begin{tabular}{cllccrcccccc}
\hline\hline
Obs. code & Run & Tstart & Exposure & \multicolumn{2}{c}{Rate } & HXD & \multicolumn{1}{c}{HXD$_1$} & HXD$_2$ & N burst & $T_{rec}$ & Type \\
& & (s) & (s) & \multicolumn{2}{c}{ (ct s$^{-1}$)} & (s) & \multicolumn{1}{c} {(s)} & (s) & (s) & \\
& & & & MECS & PDS & CCF &  &  \\
\hline
21226001  & A2b & 8994.0   & 2191.0 & 203.1 & 39.6 & 3.74 & 4.29  & 3.83 & 43 & 49.0   & S2 \\
          & A3  & 14727.0  & 2509.5 & 203.9 & 39.2 & 3.14 & 3.80  & 3.40 & 53 & 46.0   & S2 \\
          & A4  & 20648.0  & 3152.0 & 198.0 & 36.5 & 2.92 & 3.56  & 2.62 & 66 & 46.8   & M1 \\
          & A5  & 26511.0  & 2809.5 & 193.4 & 35.1 & 2.72 & 3.20  & 2.45 & 57 & [46.5] & M0 \\
          & A6  & 31928.5  & 3108.0 & 191.8 & 35.4 & 2.55 & 3.37  & 2.54 & 64 & [47.0] & M1 \\
          & A7  & 37662.0  & 3156.5 & 199.2 & 38.4 & 3.52 & 3.81  & 3.58 & 63 & [49.4] & M1 \\
          & A8b & 43402.0  & 2507.0 & 197.2 & 38.4 & 3.51 & 4.05  & 3.42 & 62 & 48.8   & S2 \\
          & A9  & 49296.0  & 2952.5 & 206.1 & 41.9 & 3.90 & 4.77  & 3.83 & 56 & 51.4   & S2 \\
212260011 & B2b & 94999.0  & 2197.0 & 207.6 & 40.6 & 3.44 & 3.92  & 3.56 & 45 & 48.0   & S1 \\
          & B3  & 100740.0 & 2638.5 & 199.6 & 38.0 & 3.07 & 3.57  & 3.07 & 55 & 47.5   & T1 \\
          & B4  & 107594.0 & 2008.0 & 207.6 & 40.5 & 3.87 & 4.56  & 3.44 & 38 & 47.0   & T1 \\
          & B5  & 112199.0 & 3093.0 & 212.2 & 42.5 & 3.84 & 4.38  & 4.06 & 61 & 49.1   & M1 \\
          & B6  & 117933.0 & 3146.5 & 208.3 & 40.7 & 3.59 & 4.15  & 4.00 & 65 & 48.0   & T2 \\
          & B7  & 123666.5 & 3122.0 & 199.3 & 38.6 & 3.34 & 3.84  & 3.12 & 62 & 49.0   & M1 \\
          & B8  & 129400.0 & 3057.5 & 203.3 & 39.2 & 3.44 & 3.90  & 3.26 & 61 & 48.5   & M1 \\
          & B9b & 136123.5 & 2130.0 & 205.8 & 41.0 & 3.57 & 3.88  & 3.33 & 42 & 45.0   & T1 \\
          & B10 & 141260.0 & 2685.0 & 210.1 & 41.6 & 3.96 & 4.52  & 3.64 & 52 & [49.7] & M1 \\
212260012 & C1  & 181003.0 & 2198.0 & 203.4 & 36.5 & 2.69 & 3.15  & 2.89 & 34 & [65.4] & M0 \\
          & C2  & 186736.5 & 2679.0 & 206.1 & 41.1 & 4.04 & 4.66  & 3.10 & 49 & 56.0   & T1 \\
          & C3  & 192470.0 & 3103.5 & 206.6 & 39.2 & 3.12 & 3.72  & 2.85 & 61 & [49.0] & M1 \\
          & C4  & 198203.5 & 3121.0 & 207.3 & 40.8 & 3.60 & 4.18  & 3.45 & 61 & 49.5   & S1 \\
          & C5  & 203937.0 & 3086.5 & 210.9 & 43.0 & 4.07 & 4.59  & 3.95 & 59 & 49.4   & M1 \\
          & C6  & 209816.5 & 2942.0 & 211.6 & 40.2 & 3.11 & 3.72  & 2.86 & 59 & [48.2] & M1 \\
          & C7  & 215404.0 & 3067.5 & 212.5 & 39.7 & 3.00 & 3.72  & 2.34 & 60 & [50.7] & M1 \\
212260013 & D5b & 267005.0 & 2219.0 & 215.3 & 40.9 & 3.70 & 4.27  & 3.32 & 43 & 50.6   & M1 \\
          & D6  & 272748.0 & 2675.0 & 211.3 & 39.7 & 3.44 & 4.07  & 2.99 & 54 & 48.6   & T1 \\
          & D9  & 289938.0 & 3117.5 & 228.7 & 39.7 & 2.34 & 2.92  & 2.61 & 50 & [64.9] & M0 \\
          & D10 & 295671.0 & 3129.5 & 227.8 & 39.0 & 2.55 & 2.97  & 3.58 & 45 & [69.1] & M0 \\
          & D11 & 301404.5 & 2987.5 & 232.2 & 39.4 & 2.44 & 2.77  & 3.07 & 39 & 68.1   & M0 \\
          & D12 & 307138.5 & 3158.0 & 225.8 & 40.2 & 2.22 & 3.07  & 2.73 & 40 & [77.5] & M0 \\
          & D13 & 313230.0 & 2730.5 & 216.5 & 39.1 & 2.64 & 3.18  & 2.70 & 40 & [71.2] & M0 \\
212260014 & E4  & 358737.5 & 2666.5 & 221.3 & 38.9 & 2.57 & 3.34  & 3.00 & 39 & [58.0] & T1 \\
          & E5  & 364470.5 & 3063.0 & 212.9 & 37.5 & 2.52 & 2.82  & 3.18 & 47 & [67.0] & M0 \\
          & E6  & 370204.0 & 3025.5 & 204.7 & 39.6 & 3.70 & 4.14  & 3.62 & 42 & [68.2] & M0 \\
          & E7  & 375937.0 & 3112.5 & 201.9 & 40.8 & 4.87 & 5.12  & 3.61 & 49 & [64.9] & M0 \\
          & E8  & 381670.5 & 3091.0 & 212.3 & 42.7 & 4.53 & 5.20  & 3.99 & 59 & 51.0   & T2 \\
          & E10 & 393136.5 & 3062.0 & 215.8 & 44.2 & 5.70 & 5.99  & 5.57 & 55 & 53.0   & S2 \\
          & E12 & 405336.0 & 2361.5 & 235.1 & 51.2 & 6.93 & 8.49  & 6.28 & 36 & 63.0   & S2 \\
212260015 & F1  & 439021.0 & 2194.0 & 243.0 & 52.5 & 7.32 & 10.5  & 6.51 & 31 & 60.8   & M1 \\
          & F2  & 444735.5 & 2644.5 & 231.3 & 48.9 & 7.09 & 8.61  & 6.43 & 40 & 64.0   & S2 \\
          & F3  & 450468.0 & 3020.0 & 231.9 & 49.5 & 6.52 & 7.99  & 5.92 & 45 & 66.0   & S1 \\
          & F4  & 456201.0 & 3102.0 & 241.8 & 51.6 & 6.86 & 9.55  & 6.88 & 47 & 65.0   & S2 \\
          & F6  & 467752.0 & 2979.5 & 237.8 & 50.6 & 6.60 & 8.29  & 6.30 & 49 & 59.2   & M2 \\
          & F7  & 473400.5 & 3099.0 & 234.3 & 49.4 & 6.37 & 8.05  & 5.57 & 51 & 58.0   & T2 \\
          & F8  & 479133.5 & 2934.5 & 233.6 & 49.2 & 6.84 & 9.05  & 5.86 & 46 & 62.0   & S2 \\
          & F9  & 485200.0 & 2744.5 & 229.6 & 48.7 & 6.64 & 8.29  & 6.09 & 44 & 61.1   & M1 \\
          & F17 & 530768.0 & 2578.0 & 244.3 & 51.9 & 7.04 & 10.9  & 5.31 & 36 & 67.0   & T1 \\
212260016 & G3  & 547929.0 & 3070.5 & 243.0 & 52.1 & 7.26 & 10.0  & 5.95 & 46 & 65.1   & M1 \\
          & G4  & 553723.5 & 3026.0 & 234.1 & 48.4 & 6.48 & 9.03  & 5.11 & 50 & 58.0   & S1 \\
          & G5  & 559395.0 & 3072.5 & 235.6 & 48.9 & 6.87 & 8.71  & 6.17 & 53 & 58.5   & M2 \\
          & G6b & 566000.5 & 2153.0 & 244.1 & 53.0 & 7.54 & 9.94  & 6.09 & 33 & 64.0   & S2 \\
          & G7  & 571145.0 & 2781.5 & 230.0 & 48.4 & 7.07 & 8.91  & 6.18 & 42 & [61.9] & M1 \\

\hline
\end{tabular}
\end{center}

\end{table*}

\begin{acknowledgements}
The authors thank the personnel of ASI Science Data
Center, particularly M. Capalbi, for help in retrieving \sax
archive data and the anonymous referee for his/her useful comments of the paper.
AD and TM thank Prof. A.A. Zdziarski for his useful discussion on the analysis. 
\\
This work has been partially supported by research
funds of the Sapienza Universit\`a di Roma. 
\end{acknowledgements}

\bibliographystyle{aa}

\bibliography{grs1915}

\end{document}